\documentclass[11pt]{article}
\usepackage[pdftex]{color}
\usepackage{graphicx}
  \usepackage{amsthm,amsfonts}
  \usepackage{amsmath}
\usepackage{slashed}
\usepackage{ulem}
\usepackage{hyperref}
\usepackage{cleveref}
\usepackage{cite}
\newcommand{\bea}   {\begin{eqnarray}}
\newcommand{\eea}   {\end{eqnarray}}

\begin{document}
\renewcommand{\thefootnote}{\fnsymbol{footnote}}

\thispagestyle{empty}

\title{Classification of minimal $ {\mathbb Z}_2\times{\mathbb Z}_2$-graded Lie \\(super)algebras and some applications}

\author{Zhanna Kuznetsova\thanks{{E-mail: {\it zhanna.kuznetsova@ufabc.edu.br}}}\quad and\quad
Francesco Toppan\thanks{{E-mail: {\it toppan@cbpf.br}}}
\\
\\
}
\maketitle

\centerline{$^{\ast}$ {\it UFABC, Av. dos Estados 5001, Bangu,}}\centerline{\it { cep 09210-580, Santo Andr\'e (SP), Brazil. }}
~\\
\centerline{$^{\dag}$
{\it CBPF, Rua Dr. Xavier Sigaud 150, Urca,}}
\centerline{\it{
cep 22290-180, Rio de Janeiro (RJ), Brazil.}}
~\\
\maketitle
\begin{abstract}
This paper presents the classification, over the fields of real and complex numbers, of the minimal  ${\mathbb Z}_2\times{\mathbb Z}_2$-graded Lie algebras and Lie superalgebras spanned by $4$ generators and with no empty graded sector. The inequivalent graded Lie (super)algebras are obtained by solving the constraints imposed by the respective graded Jacobi identities.
A motivation  for this mathematical result is to systematically investigate the properties of dynamical systems invariant under
graded (super)algebras. Recent works only paid attention to the special case of the one-dimensional ${\mathbb Z}_2\times{\mathbb Z}_2$-graded Poincar\'e superalgebra. 
As applications, we  are able to extend certain constructions originally introduced for this special superalgebra to other listed ${\mathbb Z}_2\times{\mathbb Z}_2$-graded (super)algebras. We mention, in particular, the notion of ${\mathbb Z_2}\times{\mathbb Z}_2$-graded superspace and of invariant dynamical systems (both classical worldline sigma models and {{quantum Hamiltonians}}).
As a further byproduct we point out that, contrary to ${\mathbb Z}_2\times{\mathbb Z}_2$-graded superalgebras, a theory invariant under a ${\mathbb Z}_2\times{\mathbb Z}_2$-graded algebra implies the presence of ordinary bosons and three different types of exotic bosons, with exotic bosons of different types anticommuting among themselves.
~\\\end{abstract}
\vfill

\rightline{CBPF-NF-001/21}

\rightline{	arXiv:2103.04385 [math-ph]}

\newpage
\section{Introduction}

${\mathbb Z}_2\times{\mathbb Z}_2$-graded Lie algebras and superalgebras were introduced in \cite{{rw1},{rw2}} by taking the construction of ordinary Lie superalgebras as a starting point; some possible physical applications were suggested. 
The difference between ${\mathbb Z}_2\times{\mathbb Z}_2$-graded Lie algebras and superalgebras is specified by a inner product which determines the graded brackets given by (anti)commutators. 
Ever since the graded superalgebras have been
widely investigated by mathematicians, see e.g. \cite{{sch},{sil},{szz},{csv},{csrt},{ais},{isv}}. \par
Early works considering physical applications of ${\mathbb Z}_2\times{\mathbb Z}_2$-graded Lie superalgebras are
\cite{{lr},{vas},{jyw},{zhe}}. It is nevertheless only recently
that ${\mathbb Z}_2\times{\mathbb Z}_2$-graded Lie superalgebras  have been  systematically investigated and applied to dynamical systems. It was recognized in \cite{{aktt1},{aktt2}} that they describe symmetries of L\'evy-Leblond equations; furthermore the following constructions have been introduced: ${\mathbb Z}_2\times{\mathbb Z}_2$-graded invariant worldline \cite{aktclass} and two-dimensional \cite{brusigma} sigma models, quantum mechanics \cite{{brdu},{aktquant}}, superspace \cite{brdu2}. The role of ${\mathbb Z}_2\times{\mathbb Z}_2$-graded parastatistics has been clarified in \cite{top1}  (see also \cite{{stvdj},{tol2}} and references therein for earlier works). Up to our knowledge, so far 
${\mathbb Z}_2\times{\mathbb Z}_2$-graded Lie algebras received little or no attention in connection with dynamical systems. Indeed, the recent activity for graded (super)algebras is mostly based  \cite{{aktclass},{brdu},{aktquant},{top1}} on the special case of the so-called ${\mathbb Z}_2\times{\mathbb Z}_2$-graded one-dimensional Poincar\'e superalgebra, since this is the natural generalization of the invariant superalgebra of the supersymmetric quantum mechanics \cite{wit}. \par
The main motivation of this paper is to start a systematic investigation of the properties of dynamical systems
invariant under a broader class of ${\mathbb Z}_2\times {\mathbb Z}_2$-graded (super)algebras. Accordingly, this
work presents a mathematical part which is complemented by a discussion of some selected applications.\par
On the mathematical side this paper presents the classification of the minimal  ${\mathbb Z}_2\times{\mathbb Z}_2$-graded Lie algebras and Lie superalgebras (over ${\mathbb R}$ and ${\mathbb C}$) spanned by $4$ generators and with no empty graded sector. Two tables, ${\bf 1}$ and ${\bf 2}$, are given. They respectively list the inequivalent algebras and superalgebras. The particular case of the ${\mathbb Z}_2\times{\mathbb Z}_2$-graded one-dimensional Poincar\'e superalgebra mentioned above corresponds to the $S10_{\epsilon=1}$ entry in the classification of Table {\bf 2}.
\par
The mathematical results are the basis to discuss some applications. A generalization of the notion of
 ${\mathbb Z}_2\times{\mathbb Z}_2$-graded superspace (for both algebras and superalgebras) is given in Section {\bf 4}.
As an illustration of the method, covariant derivatives for certain listed (super)algebras are obtained.
In Section {\bf 5} the derivation of new ${\mathbb Z}_2\times{\mathbb Z}_2$-graded invariant dynamical systems
(both classical worldline sigma models and {{quantum Hamiltonians}) is presented as an example.\par
As a byproduct of our results we stress the fact that a dynamical system based on a ${\mathbb Z}_2\times{\mathbb Z}_2$-graded Lie (super)algebra presents four types of fields. In the algebra case these fields are divided into ordinary bosons and three types of exotic bosons (bosons belonging to different types mutually anticommute instead of commuting). This is a new situation which should be compared with the already recognized feature, see \cite{aktclass}, that the four types of fields associated  with a ${\mathbb Z}_2\times{\mathbb Z}_2$-graded Lie superalgebra correspond to
 ordinary bosons, exotic bosons and two types of fermions (fermions of different types mutually commute). 
Further comments about the obtained results and suggested investigations are given in the Conclusions.
 \par
The scheme of the paper is the following. The classification of minimal ${\mathbb Z}_2\times{\mathbb Z}_2$-graded Lie (super)algebras is presented in Section {\bf 2}. The construction of minimal matrix representations is discussed in Section {\bf 3}. The ${\mathbb Z}_2\times{\mathbb Z}_2$-graded
superspace for both algebras and superalgebras is introduced in Section {\bf 4}. New examples of invariant dynamical systems are introduced in Section {\bf 5}. The notion of ${\mathbb Z}_2\times{\mathbb Z}_2$-graded Lie (super)algebras is recalled in Appendix {\bf A}. Minimal graded matrices are introduced in Appendix {\bf B}. Useful minimal matrix representations are given in Appendix {\bf C}. 

\section{Classification of minimal graded Lie (super)algebras}
In this Section we present the classification of the minimal graded Lie (super)algebras, with no empty graded sector, over the fields of real and complex  numbers. The three classes of cases here considered are presented in Appendix {\bf A}. For completeness and propaedeutic reasons we start with the minimal ${\mathbb Z}_2$-graded Lie superalgebras. The core result is the presentation of the minimal ${\mathbb Z}_2\times{\mathbb Z}_2$-graded Lie algebras
and  ${\mathbb Z}_2\times{\mathbb Z}_2$-graded Lie superalgebras. A minimal graded Lie (super)algebra is spanned by one and only one non-vanishing generator in each graded sector. Therefore, the minimal ${\mathbb Z}_2$-graded Lie superalgebras are spanned by two generators, while the minimal ${\mathbb Z}_2\times{\mathbb Z}_2$-graded Lie
algebras and superalgebras are spanned by $4$ generators. The inequivalent classes of these graded Lie (super)algebras are obtained by solving the constraints imposed by the respective graded Jacobi identities, see formula (\ref{gradedjacobi}). 

\subsection{The minimal ${\mathbb Z}_2$-graded Lie superalgebras}

This case presents one generator, $H$, belonging to the ${\cal G}_0$ sector and one generator, $Q$, belonging to 
the ${\cal G}_1$ sector (see Appendix {\bf A}). The most general (anti)commutators compatible with the gradings
are
\bea
\relax [H,Q]=rQ, && \{Q,Q\} = 2s H, \quad\quad \textrm{with} \quad r,s\in {\mathbb R} \quad\textrm{or}~\quad r,s\in {\mathbb C}.
\eea
The graded Jacobi identities require the constraint
\bea\label{z2constraint}
rs&=&0
\eea
to be satisfied. This constraint  is implied by taking, in the graded Jacobi identity (\ref{gradedjacobi}), either $A=H$, $B=C=Q$ or $A=B=C=Q$. \par
Three classes of solutions are recovered:
$r=s=0$, $r=0$ with $s\neq 0$ and $r\neq 0$ with $s=0$, respectively. By suitably rescaling the generators the three inequivalent minimal ${\mathbb Z}_2$-graded Lie superalgebras are\par
~~\\
$i$) the ${\mathbb Z}_2$-graded ``abelian" superalgebra 
\bea
&[H,Q]=\{Q,Q\}=0,&
\eea
$ii$) ~the ${\cal N}=1$ one-dimensional supersymmetry algebra 
\bea &[H,Q]=0, \quad\{Q,Q\}=2H,&
\eea
$iii$) the ${\mathbb Z}_2$-graded Lie superalgebra with a Grassmann generator
\bea&[H,Q]=Q,\quad \{Q,Q\}=0.&
\eea
These three superalgebras are inequivalent for both real (${\mathbb R}$) and complex (${\mathbb C}$)
number fields. \par
The ${\mathbb Z}_2$-graded ``abelian" superalgebra $i$) enters the construction of the
simplest example of superspace, see \cite{sast}, given by two  (one even and one odd) coordinates. The superalgebra $ii$) is the simplest example of superalgebra associated with the one-dimensional Supersymmetric Quantum Mechanics \cite{wit}, where $H$ is a Hamiltonian and $Q$  is its (unique) square root supersymmetry operator. The superalgebra {\it iii}) enters the topological mechanics, with $H$ playing the role of a scaling operator,
see \cite{bht}.

\subsection{The minimal ${\mathbb Z}_2\times{\mathbb Z}_2$-graded Lie algebras}

We consider here ${\mathbb Z}_2\times{\mathbb Z}_2$-graded Lie algebras with non-empty sectors.\\
The four generators $H$, $Q_i$ ($i=1,2,3$) 
can be assigned into each graded sector as
\bea
&H\in {\cal G}_{00},\quad Q_1\in{\cal G}_{10},\quad Q_2\in{\cal G}_{01},\quad Q_3\in{\cal G}_{11}.&
\eea
Since the three sectors ${\cal G}_{11}, {\cal G}_{01}, {\cal G}_{10}$ are on equal footing (see the remark $1$ in Appendix {\bf A}) any permutation of the $Q_i$'s produce equivalent algebras.\par
 The (anti)commutators defining the algebras are
\bea\label{z2z2algstrcon}
\{Q_i,Q_j\}= d_k|\epsilon_{ijk}|Q_k, &\quad& [H,Q_i] = b_i Q_i.
\eea

The six structure constants $d_i,b_i$ ($d_i,b_i\in{\mathbb K}$, with ${\mathbb K}={\mathbb R}~\textrm{or}~{\mathbb C}$) have to be constrained to satisfy the graded Jacobi identities presented in Appendix {\bf A}, see (\ref{gradedjacobi}). In (\ref{z2z2algstrcon}) $\epsilon_{ijk}$ is the totally antisymmetric tensor normalized as $\epsilon_{123}=1$. Its modulus is taken because in the right hand side of the first set of equations (\ref{z2z2algstrcon}) the anticommutators appear;  the sum over the repeated index $k$ is understood.\par
The constraints from the graded Jacobi identities come from the triples 
$H, Q_i,Q_j$ with $i\neq j$. They are
\bea\label{z2z2lieconstraints}
&d_1(b_1-b_2-b_3)=d_2(b_2-b_3-b_1)=d_3(b_3-b_1-b_2)=0.&
\eea
We classify their inequivalent solutions.\par
By rescaling the $Q_i$'s generators through $Q_i\rightarrow \lambda_iQ_i$, a given non-vanishing $d_i$, let's say
$d_3$, can be rescaled to $1$ by setting $\lambda_3=\lambda_1\lambda_2d_3$. By plugging this result, a second non-vanishing coefficient, let's say $d_2$, is rescaled as $d_2\mapsto \lambda_1^2d_2d_3$. For ${\mathbb K}={\mathbb C}$ the second coefficient can be rescaled to $1$ by setting $\lambda_1=\frac{1}{\sqrt{d_2d_3}}$. For ${\mathbb K}={\mathbb R}$ it can be set to $\pm 1$ according to the sign of the product $d_2d_3$. For ${\mathbb K}={\mathbb R}$ and three non-vanishing constants $d_i$, at least two of them can be rescaled to $+1$, while the third one can be rescaled to $\pm 1$. By taking into account that the three sectors ${\cal G}_{11}, {\cal G}_{01}, {\cal G}_{10}$ can be mutually permuted, we arrive at the following inequivalent presentations of the
$d_i$'s structure constants:
\bea
{\textrm{For}}\quad{\mathbb K}={\mathbb C}&:& (d_1,d_2,d_3) \in\{ (0,0,0), (0,0,1), (0,1,1) , (1,1,1)\}.
\eea
\bea
{\textrm{For}}\quad{\mathbb K}={\mathbb R}&:& (d_1,d_2,d_3) \in\{ (0,0,0), (0,0,1), (0,1,1) , (0,-1,1),(1,1,1),(-1,1,1)\}.
\eea
Once fixed the $d_i$'s structure constants, the $b_i$'s structure constants are determined as follows:
\bea
(d_1,d_2,d_3)=~~(0,0,0) &\Rightarrow& \textrm{no~constraint~on}~ b_1,b_2,b_3,\nonumber\\
(d_1,d_2,d_3)=~~(0,0,1) &\Rightarrow& b_3=b_1+b_2,\nonumber\\
(d_1,d_2,d_3)=(0,\pm 1,1) &\Rightarrow& b_1=0,\quad b_2=b_3,\nonumber\\
(d_1,d_2,d_3)=(\pm 1,1,1) &\Rightarrow& b_1=b_2=b_3=0.
\eea
Taking into account the rescaling of $H$ as $H\rightarrow \lambda H$ and the permutations among the ${\cal G}_{11}, {\cal G}_{01}, {\cal G}_{10}$ sectors, the list of inequivalent, minimal,  ${\mathbb Z}_2\times{\mathbb Z}_2$-graded Lie algebras (\ref{z2z2algstrcon}) is given by the following table:

\bea\nonumber
\begin{array}{|l|c|c|c|c|c|c|}\hline &d_1&d_2&d_3&b_1&b_2&b_3\\ \hline
A1_\epsilon:&\epsilon&1&1&0&0&0\\ \hline
A2_\epsilon:&0&\epsilon&1&0&0&0\\ \hline
A3_\epsilon:&0&\epsilon&1&0&1&1\\ \hline
A4:&0&0&1&0&0&0\\ \hline
A5:&0&0&1&1&-1&0\\ \hline
A6_x:&0&0&1&\frac{1}{2}-x&\frac{1}{2}+x&1\\ \hline
A7:&0&0&0&0&0&0\\ \hline
A8_{y,z}:&0&0&0&y&z&1\\ \hline
\end{array}
\eea

{\bf Table 1:}  classification of the inequivalent minimal  ${\mathbb Z}_2\times{\mathbb Z}_2$-graded algebras over ${\mathbb R}$ with non-empty graded sectors; some of the algebras are labeled by the $\epsilon =\pm 1$ sign, while continuous classes of inequivalent real superalgebras are parametrized by  $x,y,z\in{\mathbb R}$. The restrictions are
\bea\label{restrictions1}
&\epsilon =\pm 1\quad {\textrm{and}},~ {\textrm{for}}~ x,y,z\in{\mathbb R}:\qquad x\geq 0,\qquad 0\leq|y|\leq |z|\leq 1.&
\eea

{\bf Remark 1:}
The classification of the inequivalent, minimal, ${\mathbb Z}_2\times{\mathbb Z}_2$-graded Lie algebras over ${\mathbb C}$ with non-empty graded sectors are directly read from table ${\bf 1}$ by taking into account:\\
{\it i}) the irrelevance of the $\epsilon$ sign (two complex algebras differing by this sign are identified);\\
{\it ii}) that $x,y,z \in {\mathbb C}$, with  the constraint on $x$
being replaced by $0\leq\theta<\pi$ for $x=\rho e^{i\theta}$.

\subsection{The minimal ${\mathbb Z}_2\times{\mathbb Z}_2$-graded Lie superalgebras}

We consider here ${\mathbb Z}_2\times{\mathbb Z}_2$-graded Lie superalgebras with non-empty sectors.\\
The four generators $H$, $Q_1$, $Q_2$, $Z$ 
are assigned into each graded sector as
\bea
&H\in {\cal G}_{00},\quad Q_1\in{\cal G}_{10},\quad Q_2\in{\cal G}_{01},\quad Z\in{\cal G}_{11}.&
\eea
 The (anti)commutators defining these superalgebras are
\bea \label{z2z2superalgstrcon}\begin{array}{ccccccccccc}
\relax [H,Q_i]&=&a_iQ_i, && [H,Z]& =& bZ,&&[Q_1,Q_2]&=&cZ,\\
\{Q_i,Q_i\}&=&\alpha_iH, && \{Z,Q_i\} &=&\beta_i|\epsilon_{ij}|Q_j.&&&&\\
\end{array}&&
\eea
In the last equation the sum over the repeated index $j$ is understood (the antisymmetric tensor $\epsilon_{ij}$ is normalized as $\epsilon_{12}=1$).\par
The structure constants $a_1,a_2,b,c, \alpha_1,\alpha_2,\beta_1,\beta_2\in{\mathbb K}$ are constrained by the graded Jacobi identities. The complete set of nontrivial constraints derived for each choice of the $A,B,C$ generators entering equation (\ref{gradedjacobi}) is given by
\bea \label{constraints}
\begin{array}{lllcccc}
A=H,&B= Q_1,& C= Q_2&:& \quad ~c(b-a_1-a_2)&=&0,\\
A=H,&B= Q_1,& C= Z&:& \quad \beta_1(a_2-a_1-b)&=&0,\\
A=H,& B= Q_2,& C= Z&:& \quad \beta_2(a_1-a_2-b)&=&0,\\
A=Q_1,&B=Q_1,& C= Q_2&:& ~~~~~\quad c\beta_1-\alpha_1a_2&=&0,\\
A= Q_2,&B=Q_2,& C= Q_1&:& ~~~~~\quad c\beta_2+\alpha_2a_1&=&0,\\
A=Q_1,&B=Q_1,& C= Z&:& ~~~~~\quad c\beta_1-\alpha_1b&=&0,\\
A=Q_2,&B=Q_2,& C= Z&:& ~~~~~\quad c\beta_2+\alpha_2b&=&0,\\
A=Q_1,&B= Q_2,& C= Z&:&~~~ ~~\quad \beta_1\alpha_2-\alpha_1\beta_2&=&0.
\end{array}&&
\eea
By taking into account the normalization of the generators and the admissible ${\cal G}_{01}\leftrightarrow {\cal G}_{10}$ exchange of the graded sectors recalled in remark $2$ of Appendix {\bf A}, a convenient presentation of the inequivalent, minimal,
${\mathbb Z}_2\times{\mathbb Z}_2$-graded Lie superalgebras over ${\mathbb R}$ is expressed by the following table of structure constants satisfying the (\ref{constraints}) constraints:
\bea\nonumber
\begin{array}{|l|c|c|c|c|c|c|c|c|}\hline &a_1&a_2&b&c&\beta_1&\beta_2&\alpha_1&\alpha_2 \\ \hline
S1:&0&0&0&0&0&0&0&0\\ \hline
S2:&0&0&0&0&0&0&0&1\\ \hline
{S3_\epsilon}:&0&0&0&0&0&0&\epsilon&1\\ \hline
S4:&0&0&0&0&0&1&0&0\\ \hline
S5:&0&0&0&0&0&1&0&1\\ \hline
{S6_\epsilon}:&0&0&0&0&\epsilon&1&0&0\\ \hline
{S7_\epsilon}:&0&0&0&0&\epsilon&1&\epsilon&1\\ \hline
{S8}:&0&0&0&1&0&0&0&0\\ \hline
{S9}:&0&0&0&1&0&0&0&1\\ \hline
{S10_\epsilon}:&0&0&0&1&0&0&\epsilon&1\\ \hline
{S11}:&0&0&1&0&0&0&0&0\\ \hline
{S12}:&0&1&0&0&0&0&0&0\\ \hline
{S13_\epsilon}:&0&1&1&1&\epsilon&0&\epsilon&0\\ \hline
{S14}:&0&1&-1&0&0&1&0&0\\ \hline
{S15}:&0&1&1&0&1&0&0&0\\ \hline
{S16}:&0&1&1&1&0&0&0&0\\ \hline
{S17_x}:&0&1&x&0&0&0&0&0\\ \hline
{S18_{y,z}}:&1&y&z&0&0&0&0&0\\ \hline
{S19_x}:&1&x&1-x&0&0&1&0&0\\ \hline
{S20_\epsilon}:&1&1&0&0&1&\epsilon&0&0\\ \hline
{S21_y}:&1&y&1+y&1&0&0&0&0\\ \hline
\end{array}
\eea

{\bf Table 2:}  classification of the inequivalent minimal  ${\mathbb Z}_2\times{\mathbb Z}_2$-graded superalgebras over ${\mathbb R}$ with non-empty graded sectors; some of the superalgebras are labeled by the $\epsilon =\pm 1$ sign, while continuous classes of inequivalent real superalgebras are parametrized by  $x,y,z\in{\mathbb R}$. The restrictions are
\bea\label{restrictions2}
&\epsilon =\pm 1\quad {\textrm{and}},~ {\textrm{for}}~ x,y,z\in{\mathbb R}:\qquad x\neq 0,\qquad 0<|y|\leq 1,\qquad z\quad {\textrm{unconstrained}}.&
\eea

{\bf Remark 2:}
The classification of the inequivalent, minimal, ${\mathbb Z}_2\times{\mathbb Z}_2$-graded Lie superalgebras over ${\mathbb C}$ with non-empty graded sectors are directly read from table ${\bf 2}$ by taking into account:\\
{\it i}) the irrelevance of the $\epsilon$ sign (two complex superalgebras differing by this sign are identified);\\
{\it ii}) that $x,y,z \in {\mathbb C}$; the (\ref{restrictions2}) restrictions on $x,y$ hold in the complex case as well.

\section{Construction of minimal matrix representations}

The minimal, faithful, representations of the ${\mathbb Z}_2\times{\mathbb Z}_2$-graded Lie (super)algebras presented in Tables {\bf 1} and {\bf 2} are given by $4\times 4$ matrices whose nonvanishing real entries $m_i$'s are assigned as in formula (\ref{gradedmatrices}) presented in Appendix {\bf B}.  For each graded (super)algebra the closure of the (anti)commutators
puts restrictions on the $m_i$ values. The construction of these matrices is done by solving these constraints. For any given (super)algebra all resulting four matrices (one in each graded sector) are required to be nonzero.
For completeness and their possible usefulness in applications, the results of the computations are reported in Appendix {\bf C}. Solutions are obtained for all cases listed in Tables {\bf 1} and {\bf 2} with three exceptions: the algebra $A5$, the algebra $A6_x$ for $x\neq\frac{1}{2}$ and the superalgebra $S13_\epsilon$. It is easily shown that the consistency conditions, applied to the above three cases, imply that at least one of the four matrices is identically zero.\par
The presentation of the results in Appendix {\bf C} makes use of the freedom to rescale  to $1$, for the matrices of the graded sectors
${\cal G}_{10}, ~{\cal G}_{01},~{\cal G}_{11}$,  one of their nonvanishing entries. It is on the other hand convenient to keep as general as possible the diagonal entries
(eigenvalues) of the matrices $H\in {\cal G}_{00}$. 
Further comments about  the construction and the presentation of the solutions are given in Appendix {\bf C}.\par
It is useful to make some comments about certain special examples of graded (super)algebras and their matrix representations. In particular,
the ${\mathbb Z}_2\times {\mathbb Z}_2$-graded abelian (i.e., all their (anti)commutators are vanishing) algebra $A7$ and superalgebra $S1$ define
the respective graded superspaces discussed in the following Section {\bf 4}. Therefore, their respective matrices given in Appendix {\bf C} provide a matrix representation for the superspace coordinates. \par
Furthermore, the ${\mathbb Z}_2\times{\mathbb Z}_2$-graded structures of quaternions and split-quaternions are summarized as follows.

\subsection{${\mathbb Z}_2\times{\mathbb Z}_2$-graded structures of quaternions and split-quaternions}

Quaternions and split-quaternions (see \cite{split} for definitions and properties) admit a natural ${\mathbb Z}_2\times{\mathbb Z}_2$-graded structure. We present their connections with ${\mathbb Z}_2\times{\mathbb Z}_2$-graded algebras and superalgebras.\par
The composition law of the three imaginary quaternions $e_i$ ($i,j,k=1,2,3$) is
\bea
e_i\cdot e_j&=& -\delta_{ij}e_0+\epsilon_{ijk} e_k,
\eea
where $\epsilon_{123}=1$ is the totally antisymmetric tensor and $e_0$ is the identity. The three imaginary quaternions induce a realization of the $Cl(0,3)$ Clifford algebra, see \cite{kuto}.\par
The split-quaternions are defined by the three elements ${\tilde e_i}$ plus the identity ${\tilde e_0}$, with composition law expressed by
\bea
{\tilde e_i}\cdot{\tilde  e_j}&=& N_{ij}{\tilde e_0}+{\widetilde \epsilon_{ijk}} {\tilde e_k}.
\eea
The $3\times 3$ matrix $N_{ij}$ is diagonal: $
N_{ij}= diag(N_1,N_2,N_3)$, with $ N_1=-N_2=-N_3=-1$. The totally antisymmetric tensor of the split-quaternions is
${\widetilde \epsilon_{ijk}}=\epsilon_{ijk}N_k$. The three elements ${\tilde e_i}$ give a realization of the
Clifford algebra $Cl(2,1)$, see \cite{kuto}.
\par
A matrix presentation of the quaternions is
{\scriptsize{\bea\nonumber
&e_0=\left(\begin{array}{cccc}1&0&0&0\\0&1&0&0\\0&0&1&0\\0&0&0&1 \end{array}\right),\quad
e_1=\left(\begin{array}{cccc}0&0&1&0\\0&0&0&1\\ -1&0&0&0\\0&-1&0&0 \end{array}\right),\quad
e_2=\left(\begin{array}{cccc}0&0&0&1\\0&0&-1&0\\0&1&0&0\\-1&0&0&0 \end{array}\right),\quad
e_3=\left(\begin{array}{cccc}0&1&0&0\\-1&0&0&0\\0&0&0&-1\\0&0&1&0 \end{array}\right).&
\eea
}}
\bea\label{quat}
&&
\eea
A matrix presentation of the split-quaternions is
{\scriptsize{\bea\nonumber
&{\tilde e_0}=\left(\begin{array}{cccc}1&0&0&0\\0&1&0&0\\0&0&1&0\\0&0&0&1 \end{array}\right),\quad
{\tilde e_1}=\left(\begin{array}{cccc}0&0&1&0\\0&0&0&1\\ -1&0&0&0\\0&-1&0&0 \end{array}\right),\quad
{\tilde e_2}=\left(\begin{array}{cccc}0&0&0&1\\0&0&1&0\\0&1&0&0\\1&0&0&0 \end{array}\right),\quad
{\tilde e_3}=\left(\begin{array}{cccc}0&1&0&0\\1&0&0&0\\0&0&0&-1\\0&0&-1&0 \end{array}\right).&
\eea
}}
\bea\label{splitquat}
&&
\eea

Quaternionic and split-quaternionic matrices give representations of the ${\mathbb Z}_2\times {\mathbb Z}_2$-graded abelian algebra $A7$. They are recovered from the formulas given in Appendix {\bf C} for that case after setting $p=-q=-1$ (and, respectively, $p=q=-1$).\par
The quaternionic matrices (\ref{quat}) give a representation of the ${\mathbb Z}_2\times{\mathbb Z}_2$-graded superalgebra $S10_{\epsilon=1}$. They are recovered, up to normalizing factors, from the matrices presented in Appendix {\bf C} after setting,
for $\epsilon=1$, $\lambda=-2$, $p=1$, $q=-1$. \par
The split-quaternionic matrices (\ref{splitquat}) give a representation of the ${\mathbb Z}_2\times{\mathbb Z}_2$-graded superalgebra $S10_{\epsilon=-1}$. They are recovered, up to normalizing factors, from the matrices given in Appendix {\bf C} after setting,
for $\epsilon=-1$, $\lambda=2$, $p=1$, $q=1$. \par
We recall that the ${\mathbb Z}_2\times{\mathbb Z}_2$-graded superalgebra $S10_{\epsilon=1}$ was investigated in
\cite{{aktclass},{brdu},{aktquant}}. Its role, in those works, is of the one-dimensional ${\mathbb Z}_2\times {\mathbb Z}_2$-graded Poincar\'e superalgebra.

\section{Applications to generalized ${\mathbb Z}_2\times{\mathbb Z}_2$-graded superspaces}

We extend here the notion of ${\mathbb Z}_2\times{\mathbb Z}_2$-graded superspace and covariant derivatives,
originally introduced in \cite{brdu2} for the one-dimensional ${\mathbb Z}_2\times{\mathbb Z}_2$-graded Poincar\'e superalgebra $S10_{\epsilon=1}$, to other cases of graded algebras and superalgebras listed in Tables {\bf 1} and {\bf 2}, respectively.\par
At first the general construction is presented in parallel for both graded algebras and superalgebras. Then, as an illustration, after recalling the $S10_{\epsilon=1}$ superspace formulation, we present the formulas of covariant derivatives in two selected graded algebra cases, $A4$ and $A8_{y,z}$ at $y=z=0$.

{\subsection{Superfield conventions}}

In the algebra case the ${\mathbb Z}_2\times{\mathbb Z}_2$-graded  superspace is defined by the graded coordinates $x,w_1,w_2,w_3$ (with grading assignment  $x\in{\cal G}_{00}$, $w_1\in{\cal G}_{10}$, $w_2\in {\cal G}_{01}$, $
w_3\in{\cal G}_{11}$), which satisfy the graded
abelian algebra $A7$; their (anti)commutators are
\bea\label{a7anticomm}
\relax& [x,w_1]=[x,w_2]=[x,w_3]= \{w_1,w_2\}=\{w_2,w_3\}=\{w_3,w_1\}=0.&
\eea
The grading assignment of the derivative operators  is:  $\partial_x\in{\cal G}_{00}$, $ \partial_{w_1}\in{\cal G}_{10}$, $\partial_{w_2}\in {\cal G}_{01}$, $
\partial_{w_3}\in{\cal G}_{11}$. \\ The superspace coordinates and their derivatives satisfy a ${\mathbb Z}_2\times{\mathbb Z}_2$-graded algebra, whose nonvanishing (anti)commutators are
{{\bea
\relax &  [\partial_x,x]=1, \qquad ~~[\partial_{w_i},w_i ] = 1\quad {\textrm{for}}~~  i=1,2,3,\qquad ~~\{\partial_{w_i},w_j\}=\delta_{ij}\quad {\textrm{for}}~~ i\neq j.&
\eea}}
~\par
In the superalgebra case the ${\mathbb Z}_2\times{\mathbb Z}_2$-graded  superspace is defined by the graded coordinates $x,\theta,\eta,s$ (with grading assignment  $x\in{\cal G}_{00}$, $\theta\in{\cal G}_{10}$, $\eta\in {\cal G}_{01}$, $
s\in{\cal G}_{11}$), which satisfy the graded
abelian superalgebra $S1$; their (anti)commutators are
\bea
\relax& [x,\theta]=[x,\eta]=[x,s]=[\theta,\eta]= \{\theta,\theta\}=\{\eta,\eta\}=\{\theta,s\}=\{\eta,s\}=0.&\nonumber\\
\eea
The grading assignment of the derivative operators  is:  $\partial_x\in{\cal G}_{00}$, $ \partial_\theta\in{\cal G}_{10}$, $\partial_ \eta\in {\cal G}_{01}$, $
\partial_s\in{\cal G}_{11}$. \\ The superspace coordinates and their derivatives satisfy a ${\mathbb Z}_2\times{\mathbb Z}_2$-graded superalgebra, whose {{ (anti)commutators are
\bea
\relax &  [\partial_x,x]=[\partial_s,s]=\{\partial_\theta,\theta\}=\{\partial_\eta,\eta\}=1,&
\nonumber\\
\relax &[\partial_\theta,\eta]=[\partial_\eta,\theta]=\{\partial_\theta,s\}=\{\partial_s,\theta\}=\{\partial_\eta,s\}=
\{\partial_s,\eta\}=0.&
\eea 
}}

{{A ${\mathbb Z}_2\times{\mathbb Z}_2$-graded Lie algebra superfield  ${\tilde\Psi}\equiv {\tilde\Psi}(x,w_1,w_2,w_3)$ is a function of the coordinates $x,w_1,w_2,w_3$, while a ${\mathbb Z}_2\times{\mathbb Z}_2$-graded Lie superalgebra superfield  ${\overline\Psi}\equiv {\overline\Psi}(x,s,\theta,\eta)$ is a function of the coordinates $x,s,\theta,\eta$. Due to the $\theta^2=\eta^2=0$ relations, the Taylor expansion in the fermionic coordinates $\theta,\eta$ is finite.}}\par
Let us denote:\par
{\it i}) {{a point in the {\it graded Lie algebra} superspace as ${\widetilde \Phi}$.}} We have
\bea
{\widetilde \Phi} (x,w_1,w_2,w_3)&=& x H +w_1Q_1+w_2Q_2+w_3Q_3,
\eea
with $H,Q_1,Q_2,Q_3$ the generators of one of the algebras listed in Table {\bf 1};\par
{\it ii}) {{a point in the {\it graded Lie superalgebra} superspace as ${\overline \Phi}$.}}  We have
\bea
{\overline \Phi} (x,\theta,\eta,s)&=& x H +\theta Q_{10}+\eta Q_{01}+s Z,
\eea
with $H,Q_{10},Q_{01},Z$ the generators of one of the superalgebras listed in Table {\bf 2}.
\par
~\par
 The graded superspace coordinates are compactly denoted, in the algebra case, as \\${\widetilde X} =x,w_1,w_2,w_3$
and,
in the superalgebra case, as ${\overline X} =x, \theta,\eta,s$.
In order to have a unified notation we further set, depending on the case, $X\equiv {\widetilde X}, {\overline X}$ and $\Phi(X)\equiv {\widetilde \Phi}({\widetilde X}),{\overline\Phi}({\overline X})$.\par

Exponentiating ${\Phi}(X)$ maps the corresponding {{point in the superspace into the group element $g(X)=\exp(\Phi(X))$}}. \par
~\par
Different $A,B$ points in the graded superspace are expressed by different graded coordinates $X^A,X^B$, whose graded components are either $x^A, w_i^A$ and $x^B,w_i^B$ or $x^A,\theta^A,\eta^A,s^A$ and $x^B,\theta^B,\eta^B,s^B$. The (anti)commutators among all graded components are vanishing. \par
The commutators at different points are expressed in terms of the graded Lie (super)algebra structure constants, given in formula (\ref{z2z2algstrcon}) for the algebra case and in formula (\ref{z2z2superalgstrcon}) for the superalgebra case.
\par
~\par
In the algebra case we get
\bea\label{tildecommutators}
\relax [{\widetilde \Phi}^A({\widetilde X}^A),{\widetilde\Phi}^B({\widetilde X}^B)]&=&
(b_1(x^Aw_1^B-w_1^Ax^B)-d_1(w_2^Aw_3^B+w_3^Aw_2^B))\cdot Q_1+\nonumber\\
&&
(b_2(x^Aw_2^B-w_2^Ax^B)-d_2(w_3^Aw_1^B+w_1^Aw_3^B))\cdot Q_2+\nonumber\\
&&
(b_3(x^Aw_3^B-w_2^Aw_1^B)-d_3(w_1^Aw_2^B+w_2^Aw_1^B))\cdot Q_3.
\eea
~\par
In the superalgebra case we get

\bea\label{overlinecommutators}
\relax [{\overline \Phi}^A({\overline X}^A),{\overline \Phi}^B({\overline X}^B)]&=&-(\alpha_1\theta^A\theta^B+\alpha_2\eta^A\eta^B)\cdot H+\nonumber\\&&
(a_1(x^A\theta^B-x^B\theta^A)-\beta_2(\eta^As^B+s^A\eta^B))\cdot Q_{10}+\nonumber\\
&&(a_2(x^A\eta^B-x^B\eta^A)-\beta_1(\theta^As^B+s^A\theta^B))\cdot Q_{01}+\nonumber\\
&&(b(x^As^B-s^Ax^B)+c(\theta^A\eta^B-\eta^A\theta^B))\cdot Z.
\eea
~\par
{\subsection{The general construction}}

Let $\Lambda\equiv {\widetilde \Lambda}, {\overline \Lambda}$ be an infinitesimal {{element of the
${\mathbb Z}_2\times{\mathbb Z}_2$-graded Lie (super)algebra}}, whose infinitesimal graded components are respectively parametrized by $\varepsilon, \delta_i$ and $\varepsilon,\nu,\rho,\sigma$:
\bea
{\widetilde\Lambda}=\varepsilon H + \delta_1Q_1+\delta_2Q_2+\delta_3Q_3, &&{\overline{\Lambda}}=\varepsilon H+\nu Q_{10}+\rho Q_{01}+\sigma Z.
\eea
The left action induced by $\Lambda$ on a group element $g(X)$ reads
\bea
g(X)= \exp(\Phi(X)) &\mapsto&g(X')= \exp(\Phi(X'))= \exp(\Lambda)\cdot\exp(\Phi(X)).
\eea
Due to the Baker-Campbell-Hausdorff formula we get
\bea\label{bchexpansion}
\Phi(X') &=& \Phi(X)+\Lambda + \sum_{n=0}^\infty c_n\Lambda^{(n)} + O(\Lambda^2),
\eea
in terms of the (linear in $\Lambda$) multiple commutators
\bea
\Lambda^{(0)} =[\Phi(X),\Lambda], \quad&& \Lambda^{{(n+1)}} = [\Phi(X),\Lambda^{(n)}].
\eea
The coefficients $c_n$ entering (\ref{bchexpansion}) belong to a subset of the Dynkin coefficients \cite{dynkin} for the Baker-Campbell-Hausdorff expansion.\par
The $\Lambda^{(0)}$ commutators and their $\Lambda^{(n)}$ iterations are read by inserting in formulas (\ref{tildecommutators}) and (\ref{overlinecommutators}) the appropriate {{expressions}}. In particular ${\widetilde \Lambda}^{(0)}$ is recovered after setting ${\widetilde\Phi}(X^A)={\widetilde \Phi}({\widetilde X})$ and ${\widetilde\Phi}(X^B)={\widetilde \Lambda}$ in (\ref{tildecommutators}), while ${\overline \Lambda}^{(0)}$ is recovered after setting ${\overline\Phi}(X^A)={\overline \Phi}({\overline X})$ and ${\overline\Phi}(X^B)={\overline \Lambda}$ in (\ref{overlinecommutators}).\par
The infinitesimal transformations $\delta{X}=X'-X$ of the graded {{components}} can be expressed in terms of the induced covariant derivatives.\par
~\par
 In the algebra case we can set
\bea
\delta({\widetilde X}) &=&(\varepsilon {\widetilde  D}_x +\delta_1{\widetilde D}_{w_1}+\delta_2{\widetilde D}_{w_2}+\delta_3{\widetilde D}_{w_3}) ({\widetilde X}).
\eea
~\par
In the superalgebra case we have
\bea
\delta({\overline X}) &=&(\varepsilon {\overline D}_x +\nu {\overline D}_{\theta}+\rho {\overline D}_{\eta}+\sigma {\overline D}_{s}) ({\overline X}).
\eea
~\par
We illustrate now the construction of covariant derivatives in three selected examples. We recover at first the \cite{brdu2} results for $S10_{\epsilon=1}$; next we present the algebra cases $A4$ and $A8_{y=0,z=0}$.  \par
~\par

{\subsection{The superalgebra $S10_{\epsilon=\pm 1}$ revisited}}
{\it Superalgebra $S10_{\epsilon=\pm 1}$}: we recover at $\epsilon=1$ the results of \cite{brdu2}.
We recall that in this case the structure constants entering (\ref{overlinecommutators}) are $a_1=a_2=b=\beta_1=\beta_2=0$, $\alpha_1=\epsilon$, $\alpha_2=c=1$.
For this choice of structure constants the series expansion in the right hand side of (\ref{bchexpansion}) terminates at $n=0$ since
\bea
{\overline\Lambda}^{(0)}&=& [{\overline\Phi}({\overline X}),{\overline\Lambda}]= -(\epsilon\theta\nu+\eta\rho)\cdot H+(\theta\rho-\eta\nu)\cdot Z,
 \quad \quad {\overline\Lambda}^{(n)}=0\quad{\textrm{for}}\quad n\geq 1.
\eea
Taking into account that the first coefficient $c_0$ in (\ref{bchexpansion}) is $c_0=-\frac{1}{2}$, the infinitesimal transformations of the graded coordinates are
\bea
&\delta(x)= \varepsilon-\frac{1}{2}\nu\epsilon\theta-\frac{1}{2}\rho\eta,\quad  \delta(\theta)= \nu,\quad \delta(\eta)=
\rho,\quad \delta(s)=\sigma-\frac{1}{2}\rho\theta+\frac{1}{2}\nu\eta.&
\eea 
The induced covariant derivatives are
\bea
& {\overline D}_x=\partial_x,\quad  {\overline D}_\theta=\partial_\theta-\frac{1}{2}\epsilon\theta\partial_x+\frac{1}{2}\eta\partial_s ,\quad  {\overline D}_\eta=\partial_\eta-\frac{1}{2}\eta\partial_x -\frac{1}{2}\theta\partial_s,\quad  {\overline D}_s=\partial_s.&
\eea
Their (anti)commutators close the ${\mathbb Z}_2\times{\mathbb Z}_2$-graded Lie superalgebra $S10_\epsilon$
(with an overall $-1$ normalization sign in front of the structure constants). The nonvanishing ones are
\bea
&\{{\overline D}_\theta,{\overline D}_\theta\}=-\epsilon{\overline D}_x,\quad 
\{{\overline D}_\eta,{\overline D}_\eta\}=-{\overline D}_x,\quad[{\overline D}_\theta,{\overline D}_\eta]=-{\overline D}_s.&
\eea
{\subsection  {The  $A4$ and $A8_{y=0,z=0}$ algebra cases}}
We set $b_1=b_2=d_1=d_2=0$ and consider at first $b_3=0$, $d_3=1$ (algebra $A4$) and then $b_3=1$, $d_3=0$ (algebra $A8_{y,z}$ at $y=z=0$).\par
~\par
{\it Algebra $A4$}: 
\bea
{\widetilde\Lambda}^{(0)}&=& [{\widetilde\Phi}({\widetilde X}),{\widetilde\Lambda}]= -(w_1\delta_2+w_2\delta_1)\cdot Q_3,
 \quad \quad {\widetilde\Lambda}^{(n)}=0\quad{\textrm{for}}\quad n\geq 1.
\eea
The infinitesimal transformations of the graded coordinates are
\bea
&\delta(x)= \varepsilon,\quad  \delta(w_1)= \delta_1,\quad \delta(w_2)=
\delta_2,\quad \delta(w_3)=\delta_3-\frac{1}{2}\delta_2w_1-\frac{1}{2}\delta_1w_2.&
\eea
The induced covariant derivatives are
\bea
& {\widetilde D}_x=\partial_x,\quad  {\widetilde D}_{w_1}=\partial_{w_1}-\frac{1}{2}w_2\partial_{w_3} ,\quad  {\widetilde D}_{w_2}=\partial_{w_2}-\frac{1}{2}w_1\partial_{w_3},\quad  {\widetilde D}_{w_3}=\partial_{w_3}.&
\eea
The unique nonvanishing (anti)commutator is
\bea
\{{\widetilde D}_{w_1},{\widetilde D}_{w_2}\}&=& -{\widetilde D}_{w_3}.
\eea
~\par
{\it Algebra $A8_{y=0,z=0}$}: 
\bea\label{casea8}
{\widetilde\Lambda}^{(0)}&=& [{\widetilde\Phi}({\widetilde X}),{\widetilde\Lambda}]= (x\delta_3-w_2\delta_1)\cdot Q_3,
 \quad \quad {\widetilde\Lambda}^{(n+1)}= x\cdot {\widetilde \Lambda}^{(n)}.
\eea
The infinitesimal transformations of the graded coordinates are computed in terms of the $c_n$ coefficients entering the Baker-Campbell-Hausdorff expansion (\ref{bchexpansion}). Instead of directly plugging these known coefficients, an alternative method, based on the Ansatz below, allows to determine them. From (\ref{bchexpansion}) and (\ref{casea8}) the infinitesimal transformations of the graded coordinates read as
\bea
&\delta(x)= \varepsilon,\quad  \delta(w_1)= \delta_1,\quad \delta(w_2)=
\delta_2,\quad \delta(w_3)=\delta_3(1+xf(x))-w_3f(x)\varepsilon&
\eea
in terms of the function
\bea\label{generating}
f(x)&=& \sum_{n=0}^\infty c_n x^n.
\eea
The induced covariant derivatives are
\bea\label{covdeva8}
& {\widetilde D}_x=\partial_x-w_3f(x)\partial_{w_3},\quad  {\widetilde D}_{w_1}=\partial_{w_1},\quad  {\widetilde D}_{w_2}=\partial_{w_2},\quad  {\widetilde D}_{w_3}=(1+xf(x))\partial_{w_3}.&
\eea
The consistency requirement, for the anticommutator, $\{{\widetilde D}_x,{\widetilde D}_{w_3}\}\propto {\widetilde D}_{w_3}$ implies that $f(x)$ should satisfy the following Riccati equation 
for a given constant $C$:
\bea\label{riccati}
xf'(x)+2f(x) +xf(x)^2-C(1+xf(x))&=&0.
\eea
Once  determined the particular  solution $f(x)=-\frac{1}{x}$, with standard method the most general solution is expressed as $f_C(x) = \frac{C}{1-\exp({-Cx})}-\frac{1}{x}$. By requiring that, at the origin, $f(0)=c_0=-\frac{1}{2}$ 
we obtain $f(x)\equiv f_{-1}(x)$ for $C=-1$. \par
Therefore $f(x)$ entering (\ref{covdeva8}) is
\bea
f(x) &=& \frac{1}{\exp{(x)}-1}-\frac{1}{x}.
\eea
This is the generating function (\ref{generating}) of the $c_n$ coefficients entering the Baker-Campbell-Hausdorff expansion (\ref{bchexpansion}).  We get $$c_0=-\frac{1}{2},~c_1=\frac{1}{12},~c_2=0,~ c_3=-\frac{1}{720},~\ldots.$$ Since the shifted function $g(x)=f(x)+\frac{1}{2}$ is odd ($g(x)+g(-x)=0$, as easily checked), the even coefficients $c_{2n}$ for positive $n$ are all vanishing ($c_{2n}=0$ for $n=1,2,3,\ldots$). \par The series of denominators of $c_1, c_3, c_5,\ldots $ corresponds to the $A060055$ series in the OEIS (On-line Encyclopedia of Integer Sequences) database; the numerators correspond to the sequence $A060054$ in the OEIS database.\par
The only nonvanishing (anti)commutator for the covariant derivatives (\ref{covdeva8}) associated with the
$A8_{y=0,z=0}$ graded algebra is given by
\bea
 [{\widetilde D}_x, {\widetilde D}_{w_3}] &=& - {\widetilde D}_{w_3}.
\eea
~\par
\section{Classical and quantum ${\mathbb Z}_2\times{\mathbb Z}_2$-graded invariant models}

Invariant models under the one-dimensional ${\mathbb Z}_2\times{\mathbb Z}_2$-graded super-Poincar\'e algebra
have been presented in \cite{aktclass} (classical worldline sigma models) and \cite{{brdu},{aktquant}} (quantum Hamiltonians).
We show in this Section that other graded Lie algebras and superalgebras listed in Tables ${\bf 1}$ and ${\bf 2}$
can be realized as dynamical symmetries. We present a general framework and illustrate it with three examples.
The selected examples are the classical worldline sigma models invariant under the graded algebra $A1_{\epsilon=1}$ and the graded superalgebra $S7_{\epsilon=1}$, {{as well as the quantum Hamiltonians invariant under $A1_{\epsilon=1}$ and $S7_{\epsilon=1}$}}.\par
Minimal ${\mathbb Z}_2\times{\mathbb Z}_2$-graded worldline sigma models depend on four time-dependent fields.
In the graded algebra case the fields are the $00$-graded ordinary boson $x(t)$ and the exotic bosons $w_1(t), w_2(t), w_3(t)$, whose respective gradings are $10$, $01$ and $11$. In the graded superalgebra case the fields are the $00$-graded ordinary boson $x(t)$, the $11$-graded exotic boson $s(t)$ and two types of fermions (parafermions),  $\theta(t)$ and $\eta(t)$, whose respective gradings are $10$ and $01$ (see Appendix {\bf A} for the grading assignments). \par
The operator $H\in{\cal G}_{00}$ is the generator of the time translations, so that $H\propto \partial_t\cdot {\mathbb I}_4$. The time coordinate $t$ can be assumed
to be either a real or a Euclidean time. The two choices are related by a Wick rotation, see \cite{aktclass}. We work here for simplicity with the Euclidean time.\par
The operators acting on the four fields and closing a ${\mathbb Z}_2\times{\mathbb Z}_2$-graded (super)algebra are differential operators. Such $D$-module representations can be derived from the $4\times 4$ real-matrix representations (as the ones given in Appendix {\bf C}), by promoting some of the real parameters to be differential operators  (for instance, by replacing $\lambda$ entering the $S7_{\epsilon=1}$ matrices by a term proportional to $\partial_t$, as in the case discussed below).\par
The differential operators $H, Q_1,Q_2,Q_3$ of a graded algebra case (and, similarly, $H,Q_{10},Q_{01},Z$ of a graded superalgebra case) act as ${\mathbb Z}_2\times{\mathbb Z}_2$-graded Leibniz derivatives. An invariant worldline sigma model is defined by the classical action ${\cal S}=\int dt {\cal L}$, with the Lagrangian ${\cal L}$ depending on the set of four fields. The ${\mathbb Z}_2\times{\mathbb Z}_2$-graded invariance requires that each one of the four operators produces a time derivative when acting on ${\cal L}$. Let $R$ be one of these operators, we therefore have
\bea
R{\cal L} &=& \partial_t(L_R ),
\eea
for some given functional $L_R$.\par
We present now the construction of invariant  classical actions for the two cases mentioned above.
\par

{\subsection{The $A1_{\epsilon=1}$ invariant worldline action}}
~\par
The $A1_{\epsilon=1}$  $D$-module representation is obtained from 
the $A1_\epsilon$ real matrices given in Appendix {\bf C} for the $\mu=\lambda$ case, by setting
$\lambda=\partial_t$ and $\epsilon=1$. We have 
{\small{\bea\label{a1operators}
H=\left(\begin{array}{cccc}\partial_t&0&0&0\\0&\partial_t&0&0\\0&0&\partial_t&0\\0&0&0&\partial_t \end{array}\right),~ &~~~&
Q_1=\frac{1}{2}\left(\begin{array}{cccc}0&0&1&0\\0&0&0&1\\1&0&0&0\\0&1&0&0 \end{array}\right),~\nonumber\\
Q_2=\frac{1}{2}\left(\begin{array}{cccc}0&0&0&1\\0&0&1&0\\0&1&0&0\\1&0&0&0 \end{array}\right),~&~~~&
Q_3=\frac{1}{2}\left(\begin{array}{cccc}0&1&0&0\\1&0&0&0\\0&0&0&1\\0&0&1&0 \end{array}\right). 
\eea 
}}
The above operators close the $A1_{\epsilon=1}$ (anti)commutators as in Table {\bf 1} conventions.\par
The induced transformations on the $x(t),~w_i(t)$ ($i=1,2,3$) graded fields are 
\bea\label{a1action}
&\begin{array}{cccc}
Hx_{~}={\dot x}_{~}, \qquad&Q_1x_{~}=\frac{1}{2}w_1,\qquad&Q_2x_{~}=\frac{1}{2}w_2,\qquad&Q_3x_{~}=\frac{1}{2}w_3,
\\
Hw_1={\dot w_1}, \qquad&Q_1w_1=\frac{1}{2}x_{~},\qquad&Q_2w_1=\frac{1}{2}w_3,\qquad&Q_3w_1=\frac{1}{2}w_2,
\\
Hw_2={\dot w_2}, \qquad&Q_1w_2=\frac{1}{2}w_3,\qquad&Q_2w_2=\frac{1}{2}x_{~},\qquad&Q_3w_2=\frac{1}{2}w_1,
\\
Hw_3={\dot w_3}, \qquad&Q_1w_3=\frac{1}{2}w_2,\qquad&Q_2w_3=\frac{1}{2}w_1,\qquad&Q_3w_3=\frac{1}{2}x_{~},
\end{array}
&\eea
with the dot denoting a time derivative (${\dot\varphi}\equiv \frac{d\varphi}{dt}$).\par
An $A1_{\epsilon=1}$-invariant action can be constructed in terms of a kinetic term $K$ and a potential term $V(u)$, where the constant kinetic term is
\bea
K&=&\frac{1}{2}({\dot x}^2-{\dot w_1}^2-{\dot w_2}^2-{\dot w_3}^2).
\eea
The potential term $V(u)$ is chosen to be a function of $u$ given by the quadratic expression
\bea
u&=& x^2-w_1^2-w_2^2-w_3^2.
\eea
The potential term is invariant under the $Q_i$'s since  $Q_i(u)=0$.\par
The  invariant worldline action is defined by the Lagrangian
\bea\label{a1sigma}
{\cal L} &=& K+V(u).
\eea
The action ${\cal S}=\int dt {\cal L}$ is invariant since
\bea
H{\cal L} = {\dot{\cal L}}, &\quad & Q_i {\cal L} = 0 \quad {\textrm{for}}~~ i=1,2,3.
\eea
The Lagrangian ${\cal L}$ is, by construction, 00-graded due to the mod $2$ additivity (\ref{degree}).\par
The sigma model defined by (\ref{a1sigma}) is the simplest example of a non-trivial dynamical system invariant under
a ${\mathbb Z}_2\times{\mathbb Z}_2$-graded Lie algebra.  It implies the consistency of the equations of motion
involving three mutually anticommuting exotic bosons $w_i(t)$ such that $\{w_i(t),w_j(t)\}=0$ for $i\neq j$.

~\par

{{{\subsection{The $A1_{\epsilon=1}$ quantum model }}}}

{{Following the quantization prescription of \cite{aktquant} (adapted to ${\mathbb Z}_2\times{\mathbb Z}_2$-graded {\it Lie algebras}), the three graded classical fields $w_i$ are replaced by matrices $W_i$ and the classical brackets are replaced by (anti)commutators. We can set $W_i= y_i\cdot R_i$, where the $y_i$'s are real parameters and the $R_i$'s are $4\times 4$ matrices. For $i\neq j$ the anticommutators read as
$\{W_i, W_j\}=0\Rightarrow \{R_i, R_j\}=0 $. This leaves an ambiguity on the normalization sign $\sigma=\pm 1$,
introduced through $R_i^2=\sigma\cdot{\mathbb I}_4$. \\
For $\sigma=-1$ the $R_i$'s are the imaginary quaternions. For $\sigma=+1$
they are the gamma matrices of the $Cl(3,0)$ Clifford algebra. The sign ambiguity is solved by requiring the
quantum Hamiltonian, besides being Hermitian, to be bounded from below. This implies choosing $\sigma=-1$. By setting $x=y_0\cdot {\mathbb I}_4$, with $y_0$ real, the quantum potential $V$ reads as
$V(y_0^2+y_1^2+y_2^2+y_3^2)\cdot {\mathbb I}_4$. A similar analysis holds for the momenta derived by the kinetic term.}}
\par
{{Once understood the structure, we can simply write the quantum Hamiltonian $H_Q$ and check the consistency of the theory. $H_Q$ is the diagonal differential operator
\bea
&H_Q = H_q\cdot {\mathbb I}_4,\qquad {\textrm{with}} \quad H_q = -\frac{1}{2}(\partial_{y_0}^2+\partial_{y_1}^2+\partial_{y_2}^2+\partial_{y_3}^2)+V(y_0^2+y_1^2+y_2^2+y_3^2).&
\eea 
The $4\times4$ matrices $Q_1,Q_2,Q_3$ introduced in (\ref{a1operators}) are the invariant operators which, together with $H_Q$, close the quantum version of the $A1_{\epsilon=1}$ algebra.}}\par
{{The special choice of potential, $V= \frac{1}{2}(y_0^2+y_1^2+y_2^2+y_3^2)$,  gives as a quantum model $4$ copies of the $4$-dimensional harmonic oscillator.}}\par
{{{It is worth mentioning that the ${\mathbb Z}_2\times {\mathbb Z}_2$-graded invariance does not imply any new physics for the $4\times 4$ single-particle quantum Hamiltonian. Its physical consequences appear in the multiparticle sector, due to the implementation of the parastatistics associated with ${\mathbb Z}_2\times {\mathbb Z}_2$ parabosons. Following the lines of \cite{top1}, the consequences of ${\mathbb Z}_2\times {\mathbb Z}_2$ parabosons (for a different type of model which is invariant under the algebra $A7$) have been recently presented in \cite{top2}.
 }}\par
~\par

{\subsection{The $S7_{\epsilon=1}$ invariant worldline sigma model}}

~\par
A convenient $D$-module representation for the $S7_{\epsilon=1}$ superalgebra is given by the differential operators
{\small{\bea\label{dmods7}
\qquad H=\left(\begin{array}{cccc}\partial_t&0&0&0\\0&\partial_t&0&0\\0&0&\partial_t&0\\0&0&0&\partial_t \end{array}\right),~ ~~&\quad&
~~Z=\left(\begin{array}{cccc}0&\cos(\gamma)^2&0&0\\\cos(\gamma)^2 &0&0&0\\0&0&0&\sin(\gamma)^2\\0&0&\sin(\gamma)^2&0 \end{array}\right) ,\nonumber \\ 
Q_{10}=\left(\begin{array}{cccc}0&0&1&0\\0&0&0&1\\ \partial_t&0&0&0\\0&\partial_t&0&0 \end{array}\right),
~~~&\quad&
Q_{01}=\left(\begin{array}{cccc}0&0&0&1\\0&0&1&0\\0&\partial_t&0&0\\ \partial_t&0&0&0 \end{array}\right),
\eea}}
where $\gamma$ is an arbitrary angle. \\
The above representation can be recovered, at $\gamma=0$, from
the second set of $S7_{\epsilon}$ real matrices given in Appendix ${\bf C}$ by setting $\epsilon=1$, $\lambda=2\partial_t$ and by taking the special value $p=1$ (for convenience the operator $H$ has been renormalized by a $\frac{1}{2}$ factor). \par
The ${\mathbb Z}_2\times{\mathbb Z}_2$-graded (anti)commutators obtained from (\ref{dmods7}) are
\bea\label{s7superalg}
\relax [H,Z]=[H, Q_{10}]=[H,Q_{01}]=0, && [Q_{10},Q_{01}]=0,\nonumber\\
\{Q_{10},Q_{10}\}=\{Q_{01},Q_{01}\}=2H,~&& \{Z,Q_{10}\}=Q_{01},~~~~\{Z,Q_{01}\}=Q_{10}.
\eea
Since the commutator between $Q_{10}, Q_{01}$ is vanishing, the subalgebra generated by $H, Q_{10}, Q_{01}$
is the Beckers-Debergh algebra, see \cite{bede}. The presence of the extra generator $Z$ makes it to be the $S7_{\epsilon=1}$ superalgebra.\par
The action of the operators (\ref{dmods7}) on the bosonic $x(t), s(t)$ and parafermionic $\theta(t),\eta(t)$ graded fields is
\bea\label{s7action}
&\begin{array}{cccc}
Hx={\dot x}, \qquad&Zx=\cos(\gamma)^2 s,\qquad&Q_{10}x=\theta,\qquad&Q_{01}x=\eta
\\
Hs={\dot s}, \qquad&Zs=\cos(\gamma)^2 x,\qquad&Q_{10}s=\eta,\qquad&Q_{01}s=\theta,
\\
H\theta={\dot \theta}, \qquad&Z\theta=\sin(\gamma)^2\eta,\qquad&Q_{10}\theta={\dot x},\qquad&Q_{01}\theta={\dot s},
\\
H\eta={\dot \eta}, \qquad&Z\eta=\sin(\gamma)^2\theta,\qquad&Q_{10}\eta={\dot s},\qquad&Q_{01}\eta={\dot x}.
\end{array}
&\eea
The (anti)commutators of the graded fields are
\bea\label{s7anticomm}
\relax &[x(t),s(t)]=[x(t),\theta(t)]=[x(t),\eta(t)]=[\theta(t),\eta(t)]=0,&\nonumber\\
&\{\theta(t),\theta(t)\}=\{\eta(t),\eta(t)\}=\{s(t),\theta(t)\}=\{s(t),\eta(t)\}=0.&
\eea
Let us now introduce the $00$-graded quadratic expression $z$, given by
\bea
z&=& x^2-s^2.
\eea
A $S7_{\epsilon=1}$-invariant action ${\cal S}=\int dt {\cal L}$ is recovered by a Lagrangian ${\cal L}$ expressed as
\bea
{\cal L} &=& Q_{10}\cdot Q_{01}\left( f(z)\theta\eta\right),
\eea
where $f(z)$ is an arbitrary prepotential.
By construction the Lagrangian ${\cal L}$ is $00$-graded.\par
The invariance of the action ${\cal S}$ under the (\ref{dmods7}) operators, which act as graded Leibniz derivatives,
immediately follows from the fact that ${\cal L}$  transforms at most as a time derivative.
For instance, the invariance under $Z$ is implied by the identities $Z(z)=Z(\theta\eta)=0$. \par
The expression of the Lagrangian is
\bea\label{lagrangian}
{\cal L} &=& f(z)\left({\dot x}^2 -{\dot s}^2+\theta{\dot \theta}-\eta{\dot \eta}\right) +2f_z(z)\left(\theta\eta(s{\dot x}-x{\dot s})\right),
\eea
where $f_z(z)\equiv \frac{d f(z)}{dz}$.\par
The first term in (\ref{lagrangian}) is a second-order kinetic term, while the second term (which is of first-order in the time derivative) has the form of a spin-orbit interaction. \par
The construction of a classical Lagrangian for ${\mathbb Z}_2\times{\mathbb Z}_2$-graded fields satisfying (\ref{s7anticomm}) follows the scheme of reference \cite{aktclass}. The difference, with respect to that paper, 
consists in producing an action which is invariant under a new ${\mathbb Z}_2\times{\mathbb Z}_2$-graded 
superalgebra ($S7_{\epsilon=1}$ instead of the Poincar\'e superalgebra $S10_{\epsilon=1}$).\par

~\par

{\subsection{The $S7_{\epsilon=1}$ quantum model }}

In principle the classical model (\ref{lagrangian}) can be quantized with the prescription discussed in \cite{aktquant}.
\par
We know that the (anti)commutators of the parafermions $\theta $ and $\eta$ are realized by constant real matrices, while space coordinates are associated with each one of the two propagating bosons $x,s$. The prepotential $f(z)$ entering 
(\ref{lagrangian}) is reabsorbed in the kinetic term, which leads to a  $2$-dimensional constant laplacian. The derivative $\partial_zf(z)$ induces extra terms which are expressed below by the complex function $g(x,y)$ entering the quantum Hamiltonian. \par
 All in all, the following quantum operators close the $S7_{\epsilon=1}$ graded superalgebra (\ref{s7superalg}). Since no confusion arises, for simplicity we keep the same name of the operators  also in the quantum case. \par
We have
{\small{\bea\label{quantums7}
 H&=&\left(\begin{array}{cccc}H_{11}&0&0&0\\0&H_{22}&0&0\\0&0&H_{33}&0\\0&0&0&H_{44} \end{array}\right),\nonumber\\
Z&=&\left(\begin{array}{cccc}0&\cos(\gamma)^2&0&0\\\cos(\gamma)^2 &0&0&0\\0&0&0&\sin(\gamma)^2\\0&0&\sin(\gamma)^2&0 \end{array}\right) ,\nonumber \\ 
Q_{10}&=&\left(\begin{array}{cccc}0&0&\partial_x-i\partial_y+g(x,y)&0\\0&0&0&\partial_x-i\partial_y+g(x,y)\\ -\partial_x-i\partial_y+g^\ast(x,y)&0&0&0\\0&-\partial_x-i\partial_y+g^\ast(x,y)&0&0 \end{array}\right),
\nonumber\\
Q_{01}&=&\left(\begin{array}{cccc}0&0&0&\partial_x-i\partial_y+g(x,y)\\0&0&\partial_x-i\partial_y+g(x,y)&0\\0&-\partial_x-i\partial_y+g^\ast(x,y)&0&0\\- \partial_x-i\partial_y+g^\ast(x,y)&0&0&0 \end{array}\right),
\nonumber\\
&&
\eea}}
where
\bea
H_{11}=H_{22}&=&-\partial_x^2-\partial_y^2+(g^\ast-g)\partial_x-i(g+g^\ast)\partial_y+gg^\ast+g_x^\ast-ig_y^\ast,\nonumber\\
H_{33}=H_{44}&=&-\partial_x^2-\partial_y^2+(g^\ast-g)\partial_x-i(g+g^\ast)\partial_y+gg^\ast-g_x-ig_y.
\eea
The space coordinates are denoted by $x,y$. The complex function $g(x,y)$ is arbitrary. By construction the above operators are Hermitian. They close the $S7_{\epsilon=1}$ graded superalgebra (\ref{s7superalg}) for any given value of the angle $\gamma$. The quantum Hamiltonian $H$ describes a two-dimensional single-particle model. This is a new
${\mathbb Z}_2\times{\mathbb Z}_2$-graded theory. Indeed, we mention that the $S10_{\epsilon=1}$-invariant  single-particle quantum Hamiltonians presented in \cite{{brdu},{aktquant}} are one-dimensional.\par
{{
Let us now specialize the function $g(x,y)$ by setting $g(x,y)=x$. Under this choice the diagonal components $H_{11},H_{22}, H_{33}, H_{44}$ of the Hamiltonian $H$ read as
\bea
&H_{\ast }=-\partial_x^2-\partial_y^2-2ix\partial_y+x^2+\rho,  \quad( \rho =1~{\textrm{for}} ~ \ast=11,22; ~  \rho=-1 ~{\textrm{for}}~ \ast =33,44).&
\eea 
One is naturally led  to look for wavefunctions $\varphi(x,y)$ with $x\in {\mathbb R}$ and $y$ compactified on the circle ($y\in {\bf S}^1$). The eigenvalue problem reads
\bea H_{\ast} \varphi_{n,m}(x,y) &=& E_{n,m}\varphi_{n,m}(x,y).
\eea 
By separating the variables so that $\varphi_{n,m}(x,y)=\varphi_n(x)\cdot e^{imy}$ with $m\in {\mathbb Z}$, we end up with
\bea
H_{\ast} \varphi_{n}(x)e^{imy} &=& (-\partial_x^2+x^2+2mx+m^2+\rho)\varphi_n(x)e^{imy}.
\eea
By shifting $x\rightarrow {\overline x}=x+m$, the eigenvalue equation for $\varphi_n({\overline x})$ reduces to
\bea
(-\partial_{\overline{x}}^2+{\overline x}^2+\rho)\varphi_n({\overline x}) &=& E_n\varphi_n({\overline x}),
\eea
that is the equation of a harmonic oscillator.
The overall result is that the spectrum of the theory is given by the spectra of the harmonic oscillators with
an infinite degeneracy due to the presence of the $m\in {\mathbb Z}$ quantum numbers.}}\par
{{
It is quite rewarding that, even in this simple model, we end up with Hamiltonians corresponding to a $2$-dimensional particle coupled to an external electromagnetic field and in the Landau gauge. These Hamiltonians enter
the description of the quantum Hall effect (see \cite{ton} for an updated review). 
The role of the anyons for the fractional quantum Hall effect is well-established. This simple application suggests that the possible role of ${\mathbb Z}_2\times {\mathbb Z}_2$-graded parastatistics deserves investigation.
}}

~\par

\section{Conclusions}

We presented the classification of the minimal ${\mathbb Z}_2\times{\mathbb Z}_2$-graded Lie algebras and superalgebras with no empty graded sector. We pointed out that,  besides the one-dimensional ${\mathbb Z}_2\times{\mathbb Z}_2$-graded Poincar\'e superalgebra already discussed in the literature, other graded (super)algebras act as symmetries of dynamical systems. We analyzed a few cases, postponing a systematic investigation to future works. \par
Among the constructions that we discussed we mention the derivation (see subsection ${\bf 4.4}$) of the graded superspace and covariant derivatives of the $A8_{y=0,z=0}$ algebra, by solving a Riccati equation and producing
a generating function for a class of Dynkin coefficients of the Baker-Campbell-Hausdorff expansion.\par
The construction of graded invariant worldline sigma models was discussed in Section ${\bf 5}$ and two such models, for the graded algebra $A1_{\epsilon=1}$ and the graded superalgebra $S7_{\epsilon=1}$, were given. 
{{Their quantum versions have also been presented. It is rewarding that the $S7_{\epsilon=1}$-invariance produces, as special case, the quantum Hamiltonian of a two-dimensional particle in external electromagnetic field and in the Landau gauge; this type of Hamiltonians enter the description of the quantum Hall effect.}}\par
The complementary results discussed in Section {\bf 3} and shown in Appendix {\bf C} are presented in order to facilitate the model-building of ${\mathbb Z}_2\times {\mathbb Z}_2$-graded invariant theories.\par
We mention that ${\mathbb Z}_2\times{\mathbb Z}_2$-graded Lie algebras so far have not been considered as symmetries of dynamical systems, since ${\mathbb Z}_2\times{\mathbb Z}_2$-graded Lie superalgebras 
were more naturally investigated as extensions of ordinary supersymmetry. This should not be the case. As a byproduct of the results in this work we point out that, while ${\mathbb Z}_2\times{\mathbb Z}_2$-graded Lie superalgebras produce models containing ordinary bosons, exotic bosons and two types of fermions which mutually 
commute, see\cite{aktclass}, similarly ${\mathbb Z}_2\times{\mathbb Z}_2$-graded Lie algebras induce models which contain ordinary bosons and three types of exotic bosons which mutually anticommute. Ordinary and exotic bosons satisfy the graded abelian algebra $A7$, whose (anti)commutators are given in (\ref{a7anticomm}). The same (anti)commutators are also satisfied, at equal-time, for the  fields of  worldline sigma models invariant under a graded algebra (see, e.g., the (\ref{a1sigma}) Lagrangian whose action is invariant under the $A1_{\epsilon=1}$ algebra).\par
We can refer, for short, to a model based on a  ${\mathbb Z}_2\times{\mathbb Z}_2$-graded superalgebra as a
{\textit{${\mathbb Z}_2\times{\mathbb Z}_2$-graded parafermionic theory}}, while a model  based on a $ {\mathbb Z}_2\times{\mathbb Z}_2$-graded algebra can be referred to as a
{\textit{${\mathbb Z}_2\times{\mathbb Z}_2$-graded parabosonic theory}}.  It was proved in 
\cite{top1} that ${\mathbb Z}_2\times{\mathbb Z}_2$-graded parafermionic multiparticle Hamiltonians lead to new testable predictions which do not have a counterpart in ordinary bosons/fermions theories. 
{{After the first version of this paper a new work \cite{top2} derived the testable consequences of ${\mathbb Z}_2\times{\mathbb Z}_2$-graded parabosons.}}

~

  \renewcommand{\theequation}{A.\arabic{equation}}
  \setcounter{equation}{0} 
\par
~\\
{\bf{\Large{Appendix A: review of ${\mathbb Z}_2$- and ${\mathbb Z}_2\times {\mathbb Z}_2$-graded Lie (super)algebras}}}
~\par
~\par
We collect here the basic properties of the ${\mathbb Z}_2$-graded Lie algebras  (usually called Lie superalgebras in the literature, see \cite{kac}) and of the
${\mathbb Z}_2\times{\mathbb Z}_2$-graded color Lie (super)algebras \cite{{rw1},{rw2}}, together with their respective graded vector spaces. To avoid unnecessary duplications of the formulae, whenever possible, the symbols that we are introducing will be unified.
In the main text we discussed three classes of Lie (super)algebras, corresponding to
\\~
\par
~~{\it i}) the ${\mathbb Z}_2$-graded Lie algebras,\par
~{\it ii}) the ${\mathbb Z}_2\times {\mathbb Z}_2$-graded Lie algebras and\par
{\it iii}) the ${\mathbb Z}_2\times {\mathbb Z_2}$-graded Lie superalgebras.\par
~\par
Each one  of the above classes of Lie (super)algebras will be defined over the field of either real (${\mathbb R}$) or
complex (${\mathbb C}$) numbers. For each application presented in the text it is specified which numerical field has
been employed.\par
The round bracket ``$(A,B)$" denotes either a commutator ($[A,B]=AB-BA$) or an anticommutator
($\{A,B\}=AB+BA$), depending on the grading of the Lie (super)algebra generators $A,B$.\par
Let us introduce now three Lie (super)algebra generators $A,B,C$ and denote their respective gradings as the
vectors ${\vec\alpha}=deg(A)$, ${\vec\beta}=deg(B)$, ${\vec\gamma}=deg(C)$. We have that\par
~\par
in case {\it i}), $\vec{\alpha}^T={\vec\alpha}$ is a $1$-component vector, such that ${\vec\alpha}=(\alpha)$, with $\alpha\in\{0,1\}$;\par
in cases {\it ii}) and {\it iii}), ${\vec\alpha}$ is a $2$-component vector (${\vec\alpha}^T=(\alpha_1,\alpha_2)$) with $\alpha_1,\alpha_2\in \{0,1\}$.\par~\par
A inner product (${\vec\alpha}\cdot {\vec\beta}\in \{0,1\}$) is defined for a given pair of ${\vec\alpha}$, ${\vec \beta}$  gradings. It is respectively given, for each of the three classes above, as
\bea\label{inner}
\textrm{case}~ {i}): ~~ {\vec\alpha}\cdot{\vec\beta} &:=& \alpha\beta\in \{0,1\},\nonumber\\
\textrm{case}~ {ii}): ~~ {\vec\alpha}\cdot{\vec\beta} &:=& \alpha_1\beta_2-\alpha_2\beta_1\in \{0,1\},\nonumber\\
\textrm{case}~ {iii}):~~  {\vec\alpha}\cdot{\vec\beta} &:=& \alpha_1\beta_1+\alpha_2\beta_2\in \{0,1\},
\eea
where the additions on the right hand sides are taken $\textrm{mod} ~2$.\par
We are now in the position to define the bracket $(A,B)$ as
\bea\label{roundbracket}
(A,B)&:=& AB -(-1)^{{\vec\alpha}\cdot{\vec \beta}}BA,
\eea
so that
\bea
(B,A) &=& (-1)^{{\vec\alpha}\cdot{\vec \beta}+1}(A,B).
\eea
The grading $deg((A,B))$ of the Lie (super)algebra generator $(A,B)$ is
\bea\label{degree}
deg((A,B)) &=& {\vec{\alpha}}+{\vec{\beta}},
\eea
where, in each of the vector components, the sums are taken ${\textrm{mod}} ~ 2$.\par
A graded Lie (super)algebra ${\cal G}$ is endowed with a $(\cdot,\cdot):{\cal G}\times{\cal G}\rightarrow {\cal G}$ bracket defined as (\ref{roundbracket}) for each $A,B$ pair of generators in ${\cal G}$ ($A,B\in{\cal G}$). {{The bracket is defined between two homogeneous elements of definite grading.}} The degree of the (anti)commutators is defined according to (\ref{degree}). A graded Lie (super)algebra is requested to satisfy, for any $A,B,C$ triple of generators of ${\cal G}$, a graded Jacobi identity. Thanks to the compact, unified notation, in each of the three above cases, the graded Jacobi identity involving $A,B,C$ can be expressed as
\bea
\label{gradedjacobi}
 (-1)^{\vec{\gamma}\cdot\vec{\alpha}}(A,(B,C))+
 (-1)^{\vec{\alpha}\cdot\vec{\beta}}(B,(C,A))+
 (-1)^{\vec{\beta}\cdot\vec{\gamma}}(C,(A,B))&=&0.
\eea

In  the case {\it i}), the ${\mathbb Z}_2$-graded Lie algebra ${\cal G}$ is decomposed into the two sectors
\bea
{\cal G} &=& {\cal G}_0\oplus {\cal G}_1\eea
of, respectively, even and odd (also known as bosonic and fermionic) generators.\par
In the cases {\it ii}) and {\it iii}) the ${\mathbb Z}_2\times {\mathbb Z}_2$-graded Lie (super)algebra
${\cal G}$ is decomposed into the four sectors
\bea
{\cal G} &=& {\cal G}_{00}\oplus {\cal G}_{01}\oplus {\cal G}_{10}\oplus {\cal G}_{11}.
\eea

The ${\mathbb Z}_2\times{\mathbb Z}_2$-grading of the $(A,B)$ bracket is read from the entries of the table
below, at the intersection of the row ($A$) with column ($B$):
\bea
&\begin{array}{|c||c|c|c|c|}\hline 
  A\backslash B  &00&10&01&11 \\  \hline\hline
00&00&10&01&11 \\ \hline
10&10&00&11&01 \\  \hline
01&01&11&00&10\\  \hline
11&11&01&10&00\\ \hline
\end{array}&
\eea

{\centerline{{\bf Table 3:}{ ${\mathbb Z}_2\times{\mathbb Z}_2$-grading of the $(A,B)$ bracket.}}}  \par
~\par
The brackets $(A,B)$ for the ${\mathbb Z}_2\times {\mathbb Z}_2$-graded Lie {\it algebras} are (anti)commutators,
according to the table
\bea
&\begin{array}{|c||c|c|c|c|}\hline 
  A\backslash B&00&10&01&11 \\  \hline\hline
00&[\cdot,\cdot]&[\cdot,\cdot]&[\cdot,\cdot]&[\cdot,\cdot]\\ \hline
10&[\cdot,\cdot]&[\cdot,\cdot]&\{\cdot,\cdot\}&\{\cdot,\cdot\} \\  \hline
01&[\cdot,\cdot]&\{\cdot,\cdot\}&[\cdot,\cdot]&\{\cdot,\cdot\} \\  \hline
11&[\cdot,\cdot]&\{\cdot,\cdot\}&\{\cdot,\cdot\}&[\cdot,\cdot] \\ \hline
\end{array}&
\eea
{\centerline{{\bf Table 4:}{ brackets $(A,B)$ of the ${\mathbb Z}_2\times {\mathbb Z}_2$-graded Lie {algebras.}}}}  \par
~\par
The brackets $(A,B)$ for the ${\mathbb Z}_2\times {\mathbb Z}_2$-graded Lie {\it superalgebras} are (anti)commutators,
according to the table
\bea
&\begin{array}{|c|c|c|c|c|}\hline 
 A\backslash B&00&10&01&11 \\  \hline
00&[\cdot,\cdot]&[\cdot,\cdot]&[\cdot,\cdot]&[\cdot,\cdot]\\ \hline
10&[\cdot,\cdot]&\{\cdot,\cdot\}&[\cdot,\cdot]&\{\cdot,\cdot\} \\  \hline
01&[\cdot,\cdot]&[\cdot,\cdot]&\{\cdot,\cdot\}&\{\cdot,\cdot\} \\  \hline
11&[\cdot,\cdot]&\{\cdot,\cdot\}&\{\cdot,\cdot\}&[\cdot,\cdot] \\ \hline
\end{array}&
\eea
{\centerline{{\bf Table 5:}{ brackets $(A,B)$ of the ${\mathbb Z}_2\times {\mathbb Z}_2$-graded Lie {superalgebras.}}}}\par
~\par
A graded vector space $V$ is a representation space for the graded Lie (super)algebra ${\cal G}$ provided that
for any $A\in {\cal G}$ the bracket (\ref{roundbracket}) is realized by (anti)commutators through the mapping $A\mapsto {\widehat A}$, where ${\widehat A}: V\rightarrow V$ is an operator acting on $V$. The graded vector space $V$
requires the vectors $v\in V$ to be associated with a grading ${\vec \nu}$ (depending on the case, it is either
${\vec \nu}=(\nu)$ and $\nu\in\{0,1\}$ or ${\vec \nu}^T=(\nu_1,\nu_2)$ with $\nu_1,\nu_2\in\{0,1\}$). Therefore,
the graded vector spaces $V$ are 
\bea\label{gradedvector}
V=V_0\oplus V_1 \quad (\textrm{for}~~ {\mathbb Z}_2)\quad &\textrm{and}&\quad
{V} = {V}_{00}\oplus {V}_{01}\oplus {V}_{10}\oplus {V}_{11}\quad (\textrm{for}~~ {\mathbb Z}_2\times{\mathbb Z}_2).
\eea

 The compatibility of the gradings for the vector space and for the Lie (super)algebra requires a vector $v'={\widehat A}v\in V$ to have grading
$\vec{\alpha}+{\vec{\nu}}$.\par
We conclude with the following remarks:
\begin{enumerate}
\item
 In the ${\mathbb Z}_2\times{\mathbb Z}_2$-graded Lie algebra case, due to the definition (\ref{roundbracket}) of the brackets in terms of the second inner product of (\ref{inner}), the sectors
${\cal G}_{10}, {\cal G}_{01}, {\cal G}_{11}$ are on equal footing: their grading assignment can be permuted under the
$S_3$ group of permutation without altering the (anti)commutators  among the graded Lie algebra generators;
\item
 In the 
${\mathbb Z}_2\times {\mathbb Z}_2$-graded Lie superalgebra case, on the other hand, the only sectors which can be switched without altering the (anti)commutators are ${\cal G}_{10}$ and ${\cal G}_{01}$ ($S_2$ permutation), while the ${\cal G}_{11}$ sector has a different role.
\end{enumerate}

  \renewcommand{\theequation}{B.\arabic{equation}}
  \setcounter{equation}{0} 
\par
~\\
{\bf{\Large{Appendix B: minimal graded matrices}}}
~\par
~\par
We present here the general features of the constant graded matrices (whose entries are either real, ${\mathbb R}$,
or complex, ${\mathbb C}$, numbers) which can be applied to provide a matrix represention of the
graded Lie (super)algebras classified in Section {\bf 2}. We further require the matrices to act on a graded vector space, see formula (\ref{gradedvector}). Therefore, the minimal matrix representations for the
minimal ${\mathbb Z}_2$-graded Lie algebras presented in Subsection {\bf 2.1} consists of $2\times 2$ matrices, while the minimal matrix representations for the minimal ${\mathbb Z}_2\times{\mathbb Z}_2$-graded
Lie (super)algebras of Subsections {\bf 2.2} and {\bf 2.3} consist of $4\times 4$ matrices.\par
For the ${\mathbb Z}_2$-grading the even (bosonic) vectors $v_B$ and the odd (fermionic) vectors $v_F$ are respectively given by
\bea
&v_B=\left(\begin{array}{c} v\\0\end{array}\right)\in V_0, \quad\quad
v_F=\left(\begin{array}{c} 0\\v\end{array}\right)\in V_1,\quad\quad (v\in{\mathbb K},~\textrm{with} ~{\mathbb K}~ \textrm{either} ~{\mathbb R}~
\textrm{or}~ {\mathbb C}).&
\eea
The ${\cal G}_0$ sector of the ${\mathbb Z}_2$-graded Lie superalgebra is given by diagonal matrices, while the ${\cal G}_1$ sector is given by block-antidiagonal matrices. In the $2\times 2$ matrix representations we have
\bea
&M_B=\left(\begin{array}{cc} d_1&0\\0&d_2\end{array}\right)\in {\cal G}_0, \quad\quad
M_F=\left(\begin{array}{cc} 0&a_1\\a_2&0\end{array}\right)\in {\cal G}_1,\quad\quad (d_1,d_2,a_1,a_2\in{\mathbb K}).&
\eea
Bosonic (fermionic) vectors are defined in terms of the $\pm 1$ eigenspaces of the Fermion Parity Operator $N_F$,
introduced through the position
\bea
N_F &:=& \left(\begin{array}{cc} 1&0\\0&-1\end{array}\right).\eea
The Fermion Parity Operator $N_F$ allows to define the projectors $P_\pm$ onto the bosonic (fermionic) vector subspaces; we have
\bea
P_\pm &=& \frac{1}{2}({\mathbb I}_2\pm N_F), \quad\quad (P_\pm^2=P_\pm, \quad P_+P_-=P_-P_+=0),
\eea
where ${\mathbb I}_2$ is the $2\times 2$ identity matrix.
The grading ($0$ or $1$) of the vector subspaces is given by the eigenvalues of the $P_-$ projector.\par
For what concerns the ${\mathbb Z}_2\times {\mathbb Z}_2$-gradings, the vector subspaces are spanned by the
vectors
{\small{\bea\label{gradedvectors}
&v_{00}=\left(\begin{array}{c} v\\0\\0\\0\end{array}\right)\in V_{00}, \quad v_{11}=\left(\begin{array}{c} 0\\v\\0\\0\end{array}\right)\in V_{11},\quad v_{10}=\left(\begin{array}{c} 0\\0\\v\\0\end{array}\right)\in V_{10},\quad
v_{01}=\left(\begin{array}{c} 0\\0\\0\\v\end{array}\right)\in V_{01},
&\nonumber\\
&&
\eea
}}
with $v\in{\mathbb K}$.\par
Under this assignment the ${\mathbb Z}_2\times {\mathbb Z}_2$-graded $4\times 4$ matrices are
{\small{\bea\label{gradedmatrices}
M_{00}=\left(\begin{array}{cccc} m_1&0&0&0\\0&m_2&0&0\\0&0&m_3&0\\0&0&0&m_4\end{array}\right)\in {\cal G}_{00},&&M_{11}=\left(\begin{array}{cccc} 0&m_5&0&0\\m_6&0&0&0\\0&0&0&m_7\\0&0&m_8&0\end{array}\right)\in {\cal G}_{11},\nonumber\\
M_{10}=\left(\begin{array}{cccc} 0&0&m_9&0\\0&0&0&m_{10}\\m_{11}&0&0&0\\0&m_{12}&0&0\end{array}\right)\in {\cal G}_{10},&&M_{01}=\left(\begin{array}{cccc} 0&0&0&m_{13}\\0&0&m_{14}&0\\0&m_{15}&0&0\\m_{16}&0&0&0\end{array}\right)\in {\cal G}_{01},\nonumber\\&&
\eea
}}
with the entries $m_1,m_2,\ldots,m_{16}\in{\mathbb K}$.\par
Two independent ``Fermion Parity Operators", $N_1, N_2$, associated with each ${\mathbb Z}_2$ component grading, can be introduced (a third ``Fermion Parity Operator", $N_3$, is the product $N_1\cdot N_2$). These operators are expressed as
\bea
&N_1 = N_F\otimes N_F, \qquad N_2= {\mathbb I}_2\otimes N_F,  \qquad (N_3=N_1\cdot N_2= N_F\otimes {\mathbb I}_2).&
\eea
They allow to define the projectors
\bea\label{z2z2projectors}
P_{1,\pm} =\frac{1}{2}({\mathbb I}_4\pm N_1) ,&&P_{2,\pm}=\frac{1}{2}({\mathbb I}_4\pm N_2),
\eea
where ${\mathbb I}_4$ is the $4\times 4$ identity matrix.\par
Let  us denote with $p_1$ ($p_2$) an eigenvalue of the projector operator $P_{1,-}$ (and, respectively, $P_{2,-}$).  The assigned ${\mathbb Z}_2\times{\mathbb Z}_2$-gradings ($00,01,10,11$) correspond to the pair $(p_1,p_2)$ of eigenvalues of the two projectors.

  \renewcommand{\theequation}{C.\arabic{equation}}
  \setcounter{equation}{0} 
\par
~\\
{\bf{\Large{Appendix C: minimal  ${\mathbb Z}_2\times{\mathbb Z}_2$-graded matrices and (super)algebras}}}
\par ~
\par
We present here a list of  $4\times 4$ minimal matrix representations of the  ${\mathbb Z}_2\times{\mathbb Z}_2$-graded algebras and superalgebras given in Tables {\bf 1} and {\bf 2}.  The main motivation to present these results is that they can be put  into use
for different physical applications (some of them have been discussed in the main text). \par
For each one of the given cases, all four matrix generators are assumed to be nonvanishing. {{In order to make the presentation not too burdensome, the generators belonging to the 
${\cal G}_{10}, {\cal G}_{01}, {\cal G}_{11}$ sectors are suitably normalized and presented up to similarity transformations. For instance, in the $S_1$ case below, the single nonvanishing entry in the first row of each one of the three matrices $Q_{10}, Q_{01}, Z$  is normalized to $1$.
 }} The parameters $\lambda, \mu, p, q$ appearing in some of the formulas given below are assumed to be real.  In some of the cases these parameters have to be restricted to be
$\neq 0$. This occurs when $p,q$ appear in the denominator (see, e.g., the operator $Q_{10}$ entering the superalgebra case $S2$). Furthermore, the requirement $\lambda\neq 0$ guarantees a nonvanishing diagonal operator $H$
(see, e.g., the algebra case $A7$) when $H$ is a multiple of the identity.
Some graded (super)algebras do not admit a minimal $4\times 4$ representation with $4$ nonvanishing generators.  
This happens for the algebra $A5$, for the algebra $A6_x$ with $x\neq \frac{1}{2}$ and for the superalgebra $S13_\epsilon$.
A restriction on the diagonal entries of the operator $H$ can lead to a more general  form for the remaining operators. The corresponding cases are also presented. In particular the $\mu=\lambda$ restriction for the algebra $A1_{\epsilon=1}$ was applied, see (\ref{a1operators}), to derive the worldline action (\ref{a1sigma}) (the normalization of the $Q_i$'s operators below is chosen to comply with the conventions used in deriving the action). \par
~\par
{\it Minimal matrix representations of the ${\mathbb Z}_2\times{\mathbb Z}_2$-graded algebras of Table {\bf 1}:}\par
~\par
{\bf Algebra $A1_\epsilon$:}
{\scriptsize{\bea\nonumber
&H=\left(\begin{array}{cccc}\lambda&0&0&0\\0&\lambda&0&0\\0&0&\mu&0\\0&0&0&\lambda \end{array}\right),~
Q_1=\left(\begin{array}{cccc}0&0&0&0\\0&0&0&1\\0&0&0&0\\0&1&0&0 \end{array}\right),~
Q_2=\left(\begin{array}{cccc}0&0&0&1\\0&0&0&0\\0&0&0&0\\\epsilon&0&0&0 \end{array}\right),~
Q_3=\left(\begin{array}{cccc}0&1&0&0\\\epsilon&0&0&0\\0&0&0&0\\0&0&0&0 \end{array}\right);&
\eea
}}

for $\mu=\lambda$ we get

{\scriptsize{\bea\nonumber
&H=\left(\begin{array}{cccc}\lambda&0&0&0\\0&\lambda&0&0\\0&0&\lambda&0\\0&0&0&\lambda \end{array}\right),~
Q_1=\frac{1}{2}\left(\begin{array}{cccc}0&0&1&0\\0&0&0&1\\1&0&0&0\\0&1&0&0 \end{array}\right),~
Q_2=\frac{1}{2}\left(\begin{array}{cccc}0&0&0&1\\0&0&\epsilon&0\\0&1&0&0\\\epsilon&0&0&0 \end{array}\right),~
Q_3=\frac{1}{2}\left(\begin{array}{cccc}0&1&0&0\\\epsilon&0&0&0\\0&0&0&1\\0&0&\epsilon&0 \end{array}\right).&
\eea
}}

{\bf Algebra $A2_\epsilon$:}

{\scriptsize{\bea\nonumber
&H=\left(\begin{array}{cccc}\lambda&0&0&0\\0&\lambda&0&0\\0&0&\mu&0\\0&0&0&\lambda\end{array}\right),~
Q_1=\left(\begin{array}{cccc}0&0&0&0\\0&0&0&1\\0&0&0&0\\0&\epsilon&0&0 \end{array}\right),~
Q_2=\left(\begin{array}{cccc}0&1&0&0\\0&0&0&0\\0&0&0&0\\0&0&0&0 \end{array}\right)
,~
Q_3=\left(\begin{array}{cccc}0&0&0&1\\0&0&0&0\\0&0&0&0\\0&0&0&0 \end{array}\right). &
\eea
}}

for $\mu=\lambda$ we get

{\scriptsize{\bea\nonumber
&H=\left(\begin{array}{cccc}\lambda&0&0&0\\0&\lambda&0&0\\0&0&\lambda&0\\0&0&0&\lambda\end{array}\right),~
Q_1=\left(\begin{array}{cccc}0&0&p&0\\0&0&0&1-pq\\\epsilon pq^2&0&0&0\\0&\epsilon(1-pq)&0&0 \end{array}\right),~
Q_2=\left(\begin{array}{cccc}0&1&0&0\\0&0&0&0\\0&0&0&q\\0&0&0&0 \end{array}\right)
,~
Q_3=\left(\begin{array}{cccc}0&0&0&1\\0&0&0&0\\0&\epsilon q&0&0\\0&0&0&0 \end{array}\right). &
\eea
}}

{\bf Algebra $A3_\epsilon$:}
{\scriptsize{\bea\nonumber
&H=\left(\begin{array}{cccc}\lambda&0&0&0\\0&\lambda-1&0&0\\0&0&\lambda&0\\0&0&0&\lambda-1 \end{array}\right),~
Q_1=\left(\begin{array}{cccc}0&0&\epsilon-p&0\\0&0&0&\epsilon p\\1-\epsilon p&0&0&0\\0&p&0&0 \end{array}\right),~
Q_2=\left(\begin{array}{cccc}0&1&0&0\\0&0&0&0\\0&0&0&\epsilon\\0&0&0&0 \end{array}\right)
,~Q_3=\left(\begin{array}{cccc}0&0&0&1\\0&0&0&0\\0&1&0&0\\0&0&0&0 \end{array}\right). &
\eea
}}

{\bf Algebra $A4$:}
{\scriptsize{\bea\nonumber
&H=\left(\begin{array}{cccc}\lambda&0&0&0\\0&\lambda&0&0\\0&0&\mu&0\\0&0&0&\lambda \end{array}\right),~
Q_1=\left(\begin{array}{cccc}0&0&0&0\\0&0&0&1\\0&0&0&0\\0&0&0&0 \end{array}\right),~
Q_2=\left(\begin{array}{cccc}0&1&0&0\\0&0&0&0\\0&0&0&0\\0&0&0&0 \end{array}\right),~
Q_3=\left(\begin{array}{cccc}0&0&0&1\\0&0&0&0\\0&0&0&0\\0&0&0&0 \end{array}\right);&
\eea
}}

for $\mu=\lambda$ we get

{\scriptsize{\bea\nonumber
&H=\left(\begin{array}{cccc}\lambda&0&0&0\\0&\lambda&0&0\\0&0&\lambda&0\\0&0&0&\lambda \end{array}\right),~
Q_1=\left(\begin{array}{cccc}0&0&0&0\\0&0&0&1\\0&0&0&0\\0&0&0&0 \end{array}\right),~
Q_2=\left(\begin{array}{cccc}0&1&0&0\\0&0&0&0\\0&0&0&p\\0&0&0&0 \end{array}\right),~
Q_3=\left(\begin{array}{cccc}0&0&0&1\\0&0&0&0\\0&0&0&0\\0&0&0&0 \end{array}\right). &
\eea
}}

{\bf Algebra $A5$:} no $4\times 4$ matrix representation.\\

{\bf Algebra $A6_x$:} no $4\times 4$ matrix representation for $x\neq \frac{1}{2}$; $A6_{x=\frac{1}{2}}$ matrix representation:
{\scriptsize{\bea\nonumber
&H=\left(\begin{array}{cccc}\lambda&0&0&0\\0&\lambda-1&0&0\\0&0&\lambda&0\\0&0&0&\lambda-1\end{array}\right),~
Q_1=\left(\begin{array}{cccc}0&0&0&0\\0&0&0&0\\1&0&0&0\\0&1&0&0 \end{array}\right),~
Q_2=\left(\begin{array}{cccc}0&1&0&0\\0&0&0&0\\0&0&0&0\\0&0&0&0 \end{array}\right),~
Q_3=\left(\begin{array}{cccc}0&0&0&0\\0&0&0&0\\0&1&0&0\\0&0&0&0 \end{array}\right). &
\eea
}}

{\bf Algebra $A7$:}
{\scriptsize{\bea\nonumber
&H=\left(\begin{array}{cccc}\lambda&0&0&0\\0&\lambda&0&0\\0&0&\lambda&0\\0&0&0&\lambda \end{array}\right),~
Q_1=\left(\begin{array}{cccc}0&0&1&0\\0&0&0&-p\\p&0&0&0\\0&-1&0&0 \end{array}\right),~
Q_2=\left(\begin{array}{cccc}0&0&0&1\\0&0&-q&0\\0&1&0&0\\-q&0&0&0 \end{array}\right),~
Q_3=\left(\begin{array}{cccc}0&1&0&0\\pq&0&0&0\\0&0&0&p\\0&0&q&0 \end{array}\right). &
\eea
}}

{\bf Algebra $A8_{y,z}$:}
{\scriptsize{\bea\nonumber
&H=\left(\begin{array}{cccc}\lambda&0&0&0\\0&\lambda-1&0&0\\0&0&\lambda-z&0\\0&0&0&\lambda-y \end{array}\right),~Q_1=\left(\begin{array}{cccc}0&0&0&1\\0&0&0&0\\0&0&0&0\\0&0&0&0 \end{array}\right)
,~
Q_2=\left(\begin{array}{cccc}0&0&1&0\\0&0&0&0\\0&0&0&0\\0&0&0&0 \end{array}\right),~
Q_3=\left(\begin{array}{cccc}0&1&0&0\\0&0&0&0\\0&0&0&0\\0&0&0&0 \end{array}\right)&
\eea
}}

and, for $y=1-z$:

{\scriptsize{\bea\nonumber
&H=\left(\begin{array}{cccc}\lambda&0&0&0\\0&\lambda-1&0&0\\0&0&\lambda-z&0\\0&0&0&\lambda-1+z \end{array}\right),~
Q_1=\left(\begin{array}{cccc}0&0&0&1\\0&0&0&0\\0&p&0&0\\0&0&0&0 \end{array}\right),~
Q_2=\left(\begin{array}{cccc}0&0&1&0\\0&0&0&0\\0&0&0&0\\0&-p&0&0 \end{array}\right),~
Q_3=\left(\begin{array}{cccc}0&1&0&0\\0&0&0&0\\0&0&0&0\\0&0&0&0 \end{array}\right).&
\eea
}}

{\it Minimal matrix representations of the ${\mathbb Z}_2\times{\mathbb Z}_2$-graded superalgebras of Table {\bf 2}:}

~\par
{\bf Superalgebra $S1$:}
{\scriptsize{\bea\nonumber
&H=\left(\begin{array}{cccc}\lambda&0&0&0\\0&\lambda&0&0\\0&0&\lambda&0\\0&0&0&\lambda \end{array}\right),~
Q_{10}=\left(\begin{array}{cccc}0&0&1&0\\0&0&0&p\\0&0&0&0\\0&0&0&0 \end{array}\right),~
Q_{01}=\left(\begin{array}{cccc}0&0&0&1\\0&0&-q&0\\0&0&0&0\\0&0&0&0 \end{array}\right),~
Z=\left(\begin{array}{cccc}0&1&0&0\\-pq&0&0&0\\0&0&0&-p\\0&0&q&0 \end{array}\right). &
\eea
}}

{\bf Superalgebra $S2$:}
{\scriptsize{\bea\nonumber
&H=\left(\begin{array}{cccc}\lambda&0&0&0\\0&\lambda&0&0\\0&0&\lambda&0\\0&0&0&\lambda \end{array}\right),~
Q_{10}=\left(\begin{array}{cccc}0&0&1&0\\0&0&0&0\\0&0&0&0\\0&-\frac{\lambda}{2p}&0&0 \end{array}\right),~
Q_{01}=\left(\begin{array}{cccc}0&0&0&1\\0&0&-p&0\\0&-\frac{\lambda}{2p}&0&0\\\frac{\lambda}{2}&0&0&0 \end{array}\right),~
Z=\left(\begin{array}{cccc}0&1&0&0\\0&0&0&0\\0&0&0&0\\0&0&p&0 \end{array}\right). &
\eea
}}

{\bf Superalgebra $S3_\epsilon$:}
{\scriptsize{\bea\nonumber
&H=\left(\begin{array}{cccc}\lambda&0&0&0\\0&\lambda&0&0\\0&0&\lambda&0\\0&0&0&\lambda \end{array}\right),~
Q_{10}=\left(\begin{array}{cccc}0&0&1&0\\0&0&0&-\epsilon p\\\frac{\epsilon\lambda}{2}&0&0&0\\0&-\frac{\lambda}{2p}&0&0 \end{array}\right),~
Q_{01}=\left(\begin{array}{cccc}0&0&0&1\\0&0&-p&0\\0&-\frac{\lambda}{2p}&0&0\\\frac{\lambda}{2}&0&0&0 \end{array}\right),~
Z=\left(\begin{array}{cccc}0&1&0&0\\ \epsilon p^2&0&0&0\\0&0&0&\epsilon p\\0&0&p&0 \end{array}\right). &
\eea
}}

{\bf Superalgebra $S4$:}
{\scriptsize{\bea\nonumber
&H=\left(\begin{array}{cccc}\lambda&0&0&0\\0&\lambda&0&0\\0&0&\lambda&0\\0&0&0&\lambda \end{array}\right),~
Q_{10}=\left(\begin{array}{cccc}0&0&1&0\\0&0&0&0\\0&0&0&0\\0&0&0&0 \end{array}\right),~
Q_{01}=\left(\begin{array}{cccc}0&0&0&1\\0&0&1-p&0\\0&0&0&0\\0&0&0&0 \end{array}\right),~
Z=\left(\begin{array}{cccc}0&1&0&0\\0&0&0&0\\0&0&0&0\\0&0&p&0 \end{array}\right). &
\eea
}}

{\bf Superalgebra $S5$:}
{\scriptsize{\bea\nonumber
&H=\left(\begin{array}{cccc}\lambda&0&0&0\\0&\lambda&0&0\\0&0&\lambda&0\\0&0&0&\lambda \end{array}\right),~
Q_{10}=\left(\begin{array}{cccc}0&0&1&0\\0&0&0&0\\0&0&0&0\\0&\frac{\lambda}{2p}&0&0 \end{array}\right),~
Q_{01}=\left(\begin{array}{cccc}0&0&0&1\\0&0&p&0\\0&\frac{\lambda}{2p}&0&0\\\frac{\lambda}{2}&0&0&0 \end{array}\right),~
Z=\left(\begin{array}{cccc}0&1&0&0\\0&0&0&0\\0&0&0&0\\0&0&1-p&0 \end{array}\right). &
\eea
}}

{\bf Superalgebra $S6_\epsilon$:}
{\scriptsize{\bea\nonumber
&H=\left(\begin{array}{cccc}\lambda&0&0&0\\0&\mu&0&0\\0&0&\lambda&0\\0&0&0&\lambda \end{array}\right),~
Q_{10}=\left(\begin{array}{cccc}0&0&1&0\\0&0&0&0\\0&0&0&0\\0&0&0&0 \end{array}\right),~
Q_{01}=\left(\begin{array}{cccc}0&0&0&1\\0&0&0&0\\0&0&0&0\\0&0&0&0 \end{array}\right),~
Z=\left(\begin{array}{cccc}0&0&0&0\\0&0&0&0\\0&0&0&\epsilon\\0&0&1&0 \end{array}\right) ;&
\eea
}}
for $\mu=\lambda$ we get
{\scriptsize{\bea\nonumber
&H=\left(\begin{array}{cccc}\lambda&0&0&0\\0&\lambda&0&0\\0&0&\lambda&0\\0&0&0&\lambda \end{array}\right),~
Q_{10}=\left(\begin{array}{cccc}0&0&1&0\\0&0&0&p\\0&0&0&0\\0&0&0&0 \end{array}\right),~
Q_{01}=\left(\begin{array}{cccc}0&0&0&1\\0&0&p&0\\0&0&0&0\\0&0&0&0 \end{array}\right),~
Z=\left(\begin{array}{cccc}0&q&0&0\\\epsilon p^2q&0&0&0\\0&0&0&\epsilon(1-pq)\\0&0&1-pq&0 \end{array}\right) .&
\eea
}}

{\bf Superalgebra $S7_\epsilon$:}
{\scriptsize{\bea\nonumber
&H=\left(\begin{array}{cccc}\lambda&0&0&0\\0&\lambda&0&0\\0&0&\lambda&0\\0&0&0&\lambda \end{array}\right),~
Q_{10}=\left(\begin{array}{cccc}0&0&1&0\\0&0&0&\epsilon p \\\frac{\epsilon}{2\lambda}&0&0&0\\0&\frac{\lambda}{2p}&0&0 \end{array}\right),~
Q_{01}=\left(\begin{array}{cccc}0&0&0&1\\0&0&p&0\\0&\frac{\lambda}{2p}&0&0\\\frac{\lambda}{2}&0&0&0 \end{array}\right),~
Z=\left(\begin{array}{cccc}0&1&0&0\\\epsilon p^2&0&0&0\\0&0&0&\epsilon (1-p)\\0&0&1-p&0 \end{array}\right) &
\eea
}}
and
{\scriptsize{\bea\nonumber
&H=\left(\begin{array}{cccc}\lambda&0&0&0\\0&\lambda&0&0\\0&0&\lambda&0\\0&0&0&\lambda \end{array}\right),~
Q_{10}=\left(\begin{array}{cccc}0&0&p&0\\0&0&0&{\epsilon}\\\frac{\epsilon \lambda}{2p}&0&0&0\\0&\frac{\lambda}{2}&0&0 \end{array}\right),~
Q_{01}=\left(\begin{array}{cccc}0&0&0&1\\0&0&p&0\\0&\frac{\lambda}{2p}&0&0\\\frac{\lambda}{2}&0&0&0 \end{array}\right),~
Z=\left(\begin{array}{cccc}0&1&0&0\\\epsilon &0&0&0\\0&0&0&0\\0&0&0&0 \end{array}\right) .&
\eea
}}

{\bf Superalgebra $S8$:}
{\scriptsize{\bea\nonumber
&H=\left(\begin{array}{cccc}\lambda&0&0&0\\0&\lambda&0&0\\0&0&\lambda&0\\0&0&0&\mu \end{array}\right),~
Q_{10}=\left(\begin{array}{cccc}0&0&1&0\\0&0&0&0 \\0&0&0&0\\0&0&0&0 \end{array}\right),~
Q_{01}=\left(\begin{array}{cccc}0&0&0&0\\0&0&0&0\\0&1&0&0\\0&0&0&0 \end{array}\right),~
Z=\left(\begin{array}{cccc}0&1&0&0\\0&0&0&0\\0&0&0&0\\0&0&0&0 \end{array}\right) &
\eea
}}
and
{\scriptsize{\bea\nonumber
&H=\left(\begin{array}{cccc}\lambda&0&0&0\\0&\lambda&0&0\\0&0&\lambda&0\\0&0&0&\lambda \end{array}\right),~
Q_{10}=\left(\begin{array}{cccc}0&0&1&0\\0&0&0&p\\0&0&0&0\\0&0&0&0 \end{array}\right),~
Q_{01}=\left(\begin{array}{cccc}0&0&0&0\\0&0&0&0\\0&1&0&0\\q&0&0&0 \end{array}\right),~
Z=\left(\begin{array}{cccc}0&1&0&0\\pq &0&0&0\\0&0&0&-p\\0&0&-q&0 \end{array}\right) .&
\eea
}}

{\bf Superalgebra $S9$:}
{\scriptsize{\bea\nonumber
&H=\left(\begin{array}{cccc}\lambda&0&0&0\\0&\lambda&0&0\\0&0&\lambda&0\\0&0&0&\lambda \end{array}\right),~
Q_{10}=\left(\begin{array}{cccc}0&0&1&0\\0&0&0&p\\0&0&0&0\\0&0&0&0 \end{array}\right),~
Q_{01}=\left(\begin{array}{cccc}0&0&0&1\\0&0&q&0\\0&\frac{\lambda}{2q}&0&0\\\frac{\lambda}{2}&0&0&0 \end{array}\right),~
Z=\left(\begin{array}{cccc}0&\frac{\lambda}{2q}&0&0\\\frac{\lambda p}{2}&0&0&0\\0&0&0&-\frac{\lambda p}{2q}\\0&0&-\frac{\lambda}{2}&0 \end{array}\right).&
\eea
}}

{\bf Superalgebra $S10_\epsilon$:}
{\scriptsize{\bea\nonumber
&H=\left(\begin{array}{cccc}\lambda&0&0&0\\0&\lambda&0&0\\0&0&\lambda&0\\0&0&0&\lambda \end{array}\right),
Q_{10}=\left(\begin{array}{cccc}0&0&1&0\\0&0&0&p\\ \frac{\epsilon\lambda}{2}&0&0&0\\0&\frac{\epsilon\lambda}{2p}&0&0 \end{array}\right),
Q_{01}=\left(\begin{array}{cccc}0&0&0&1\\0&0&q&0\\0&\frac{\lambda}{2q}&0&0\\\frac{\lambda}{2}&0&0&0 \end{array}\right),
Z=\left(\begin{array}{cccc}0&\frac{\lambda(p-\epsilon q)}{2pq}&0&0\\\frac{\lambda (p-\epsilon q)}{2}&0&0&0\\0&0&0&\frac{\lambda (\epsilon q-p)}{2q}\\0&0&\frac{\lambda(\epsilon q-p)}{2p}&0 \end{array}\right).&
\eea
}}

{\bf Superalgebra $S11$:}
{\scriptsize{\bea\nonumber
&H=\left(\begin{array}{cccc}\lambda&0&0&0\\0&\lambda-1&0&0\\0&0&\lambda&0\\0&0&0&\lambda \end{array}\right),~
Q_{10}=\left(\begin{array}{cccc}0&0&1&0\\0&0&0&0\\0&0&0&0\\0&0&0&0 \end{array}\right),~
Q_{01}=\left(\begin{array}{cccc}0&0&0&1\\0&0&0&0\\0&0&0&0\\0&0&0&0 \end{array}\right),~
Z=\left(\begin{array}{cccc}0&1&0&0\\0&0&0&0\\0&0&0&0\\0&0&0&0 \end{array}\right).&
\eea
}}

{\bf Superalgebra $S12$:}
{\scriptsize{\bea\nonumber
&H=\left(\begin{array}{cccc}\lambda&0&0&0\\0&\lambda&0&0\\0&0&\lambda&0\\0&0&0&\lambda -1\end{array}\right),~
Q_{10}=\left(\begin{array}{cccc}0&0&1&0\\0&0&0&0\\0&0&0&0\\0&0&0&0 \end{array}\right),~
Q_{01}=\left(\begin{array}{cccc}0&0&0&1\\0&0&0&0\\0&0&0&0\\0&0&0&0 \end{array}\right),~
Z=\left(\begin{array}{cccc}0&1&0&0\\0&0&0&0\\0&0&0&0\\0&0&0&0 \end{array}\right).&
\eea
}}

{\bf Superalgebra $S13_\epsilon$:} no $4\times 4$ matrix representation.\\
~\par

{\bf Superalgebra $S14$:}
{\scriptsize{\bea\nonumber
&H=\left(\begin{array}{cccc}\lambda&0&0&0\\0&\mu&0&0\\0&0&\lambda&0\\0&0&0&\lambda -1\end{array}\right),~
Q_{10}=\left(\begin{array}{cccc}0&0&1&0\\0&0&0&0\\0&0&0&0\\0&0&0&0 \end{array}\right),~
Q_{01}=\left(\begin{array}{cccc}0&0&0&1\\0&0&0&0\\0&0&0&0\\0&0&0&0 \end{array}\right),~
Z=\left(\begin{array}{cccc}0&0&0&0\\0&0&0&0\\0&0&0&0\\0&0&1&0 \end{array}\right);&
\eea
}}
for $\mu=\lambda$ we get
{\scriptsize{\bea\nonumber
&H=\left(\begin{array}{cccc}\lambda&0&0&0\\0&\lambda+1&0&0\\0&0&\lambda&0\\0&0&0&\lambda -1\end{array}\right),~
Q_{10}=\left(\begin{array}{cccc}0&0&1&0\\0&0&0&0\\0&0&0&0\\0&0&0&0 \end{array}\right),~
Q_{01}=\left(\begin{array}{cccc}0&0&0&1\\0&0&q&0\\0&0&0&0\\0&0&0&0 \end{array}\right),~
Z=\left(\begin{array}{cccc}0&p&0&0\\0&0&0&0\\0&0&0&0\\0&0&1-pq&0 \end{array}\right).&
\eea
}}

{\bf Superalgebra $S15$:}
{\scriptsize{\bea\nonumber
&H=\left(\begin{array}{cccc}\lambda&0&0&0\\0&\mu&0&0\\0&0&\lambda&0\\0&0&0&\lambda -1\end{array}\right),~
Q_{10}=\left(\begin{array}{cccc}0&0&1&0\\0&0&0&0\\0&0&0&0\\0&0&0&0 \end{array}\right),~
Q_{01}=\left(\begin{array}{cccc}0&0&0&1\\0&0&0&0\\0&0&0&0\\0&0&0&0 \end{array}\right),~
Z=\left(\begin{array}{cccc}0&0&0&0\\0&0&0&0\\0&0&0&1\\0&0&0&0 \end{array}\right);&
\eea
}}
for $\mu=\lambda -1$ we get
{\scriptsize{\bea\nonumber
&H=\left(\begin{array}{cccc}\lambda&0&0&0\\0&\lambda-1&0&0\\0&0&\lambda&0\\0&0&0&\lambda -1\end{array}\right),~
Q_{10}=\left(\begin{array}{cccc}0&0&1&0\\0&0&0&p\\0&0&0&0\\0&0&0&0 \end{array}\right),~
Q_{01}=\left(\begin{array}{cccc}0&0&0&1\\0&0&0&0\\0&0&0&0\\0&0&0&0 \end{array}\right),~
Z=\left(\begin{array}{cccc}0&q&0&0\\0&0&0&0\\0&0&0&1-pq\\0&0&0&0 \end{array}\right).&
\eea
}}

{\bf Superalgebra $S16$:}
{\scriptsize{\bea\nonumber
&H=\left(\begin{array}{cccc}\lambda&0&0&0\\0&\lambda-1&0&0\\0&0&\lambda&0\\0&0&0&\mu\end{array}\right),~
Q_{10}=\left(\begin{array}{cccc}0&0&1&0\\0&0&0&0\\0&0&0&0\\0&0&0&0 \end{array}\right),~
Q_{01}=\left(\begin{array}{cccc}0&0&0&0\\0&0&0&0\\0&1&0&0\\0&0&0&0 \end{array}\right),~
Z=\left(\begin{array}{cccc}0&1&0&0\\0&0&0&0\\0&0&0&0\\0&0&0&0 \end{array}\right);&
\eea
}}
for $\mu=\lambda+1$ we get
{\scriptsize{\bea\nonumber
&H=\left(\begin{array}{cccc}\lambda&0&0&0\\0&\lambda-1&0&0\\0&0&\lambda&0\\0&0&0&\lambda +1\end{array}\right),~
Q_{10}=\left(\begin{array}{cccc}0&0&1&0\\0&0&0&0\\0&0&0&0\\0&0&0&0 \end{array}\right),~
Q_{01}=\left(\begin{array}{cccc}0&0&0&0\\0&0&0&0\\0&1&0&0\\-p&0&0&0 \end{array}\right),~
Z=\left(\begin{array}{cccc}0&1&0&0\\0&0&0&0\\0&0&0&0\\0&0&p&0 \end{array}\right);&
\eea
}}
for $\mu=\lambda-1$ we get
{\scriptsize{\bea\nonumber
&H=\left(\begin{array}{cccc}\lambda&0&0&0\\0&\lambda-1&0&0\\0&0&\lambda&0\\0&0&0&\lambda -1\end{array}\right),~
Q_{10}=\left(\begin{array}{cccc}0&0&1&0\\0&0&0&p\\0&0&0&0\\0&0&0&0 \end{array}\right),~
Q_{01}=\left(\begin{array}{cccc}0&0&0&0\\0&0&0&0\\0&1&0&0\\0&0&0&0 \end{array}\right),~
Z=\left(\begin{array}{cccc}0&1&0&0\\0&0&0&0\\0&0&0&-p\\0&0&0&0 \end{array}\right).&
\eea
}}

{\bf Superalgebra $S17_x$:}
{\scriptsize{\bea\nonumber
&H=\left(\begin{array}{cccc}\lambda&0&0&0\\0&\lambda-x&0&0\\0&0&\lambda&0\\0&0&0&\lambda-1\end{array}\right),~
Q_{10}=\left(\begin{array}{cccc}0&0&1&0\\0&0&0&0\\0&0&0&0\\0&0&0&0 \end{array}\right),~
Q_{01}=\left(\begin{array}{cccc}0&0&0&1\\0&0&0&0\\0&0&0&0\\0&0&0&0 \end{array}\right),~
Z=\left(\begin{array}{cccc}0&1&0&0\\0&0&0&0\\0&0&0&0\\0&0&0&0 \end{array}\right).&
\eea
}}

{\bf Superalgebra $S18_{y,z}$:}
{\scriptsize{\bea\nonumber
&H=\left(\begin{array}{cccc}\lambda&0&0&0\\0&\lambda-z&0&0\\0&0&\lambda-1&0\\0&0&0&\lambda-y\end{array}\right),~
Q_{10}=\left(\begin{array}{cccc}0&0&1&0\\0&0&0&0\\0&0&0&0\\0&0&0&0 \end{array}\right),~
Q_{01}=\left(\begin{array}{cccc}0&0&0&1\\0&0&0&0\\0&0&0&0\\0&0&0&0 \end{array}\right),~
Z=\left(\begin{array}{cccc}0&1&0&0\\0&0&0&0\\0&0&0&0\\0&0&0&0 \end{array}\right);&
\eea
}}

for $z=y-1$ we get

{\scriptsize{\bea\nonumber
&H=\left(\begin{array}{cccc}\lambda&0&0&0\\0&\lambda+1-y&0&0\\0&0&\lambda-1&0\\0&0&0&\lambda-y\end{array}\right),~
Q_{10}=\left(\begin{array}{cccc}0&0&1&0\\0&0&0&p\\0&0&0&0\\0&0&0&0 \end{array}\right),~
Q_{01}=\left(\begin{array}{cccc}0&0&0&1\\0&0&0&0\\0&0&0&0\\0&0&0&0 \end{array}\right),~
Z=\left(\begin{array}{cccc}0&1&0&0\\0&0&0&0\\0&0&0&-p\\0&0&0&0 \end{array}\right);&
\eea
}}

for $z=1-y$ we get
{\scriptsize{\bea\nonumber
&H=\left(\begin{array}{cccc}\lambda&0&0&0\\0&\lambda+y-1&0&0\\0&0&\lambda-1&0\\0&0&0&\lambda-y\end{array}\right),~
Q_{10}=\left(\begin{array}{cccc}0&0&1&0\\0&0&0&0\\0&0&0&0\\0&0&0&0 \end{array}\right),~
Q_{01}=\left(\begin{array}{cccc}0&0&0&1\\0&0&-p&0\\0&0&0&0\\0&0&0&0 \end{array}\right),~
Z=\left(\begin{array}{cccc}0&1&0&0\\0&0&0&0\\0&0&0&0\\0&0&p&0 \end{array}\right);&
\eea
}}

for $z=0$, $y=1$ we get
{\scriptsize{\bea\nonumber
&H=\left(\begin{array}{cccc}\lambda&0&0&0\\0&\lambda&0&0\\0&0&\lambda-1&0\\0&0&0&\lambda-1\end{array}\right),~
Q_{10}=\left(\begin{array}{cccc}0&0&1&0\\0&0&0&p\\0&0&0&0\\0&0&0&0 \end{array}\right),~
Q_{01}=\left(\begin{array}{cccc}0&0&0&1\\0&0&-q&0\\0&0&0&0\\0&0&0&0 \end{array}\right),~
Z=\left(\begin{array}{cccc}0&1&0&0\\-pq&0&0&0\\0&0&0&-p\\0&0&q&0 \end{array}\right).&
\eea
}}

{\bf Superalgebra $S19_{x}$:}
{\scriptsize{\bea\nonumber
&H=\left(\begin{array}{cccc}\lambda&0&0&0\\0&\mu&0&0\\0&0&\lambda-1&0\\0&0&0&\lambda-x\end{array}\right),~
Q_{10}=\left(\begin{array}{cccc}0&0&1&0\\0&0&0&0\\0&0&0&0\\0&0&0&0 \end{array}\right),~
Q_{01}=\left(\begin{array}{cccc}0&0&0&1\\0&0&0&0\\0&0&0&0\\0&0&0&0 \end{array}\right),~
Z=\left(\begin{array}{cccc}0&0&0&0\\0&0&0&0\\0&0&0&0\\0&0&1&0 \end{array}\right);&
\eea
}}

for $\mu=\lambda+x-1$ we get

{\scriptsize{\bea\nonumber
&H=\left(\begin{array}{cccc}\lambda&0&0&0\\0&\lambda+x-1&0&0\\0&0&\lambda-1&0\\0&0&0&\lambda-x\end{array}\right),~
Q_{10}=\left(\begin{array}{cccc}0&0&1&0\\0&0&0&p\\0&0&0&0\\0&0&0&0 \end{array}\right),~
Q_{01}=\left(\begin{array}{cccc}0&0&0&1\\0&0&q&0\\0&0&0&0\\0&0&0&0 \end{array}\right),~
Z=\left(\begin{array}{cccc}0&p&0&0\\0&0&0&0\\0&0&0&0\\0&0&1-pq&0 \end{array}\right).&
\eea
}}

{\bf Superalgebra $S20_{\epsilon}$:}
{\scriptsize{\bea\nonumber
&H=\left(\begin{array}{cccc}\lambda&0&0&0\\0&\mu&0&0\\0&0&\lambda-1&0\\0&0&0&\lambda-1\end{array}\right),~
Q_{10}=\left(\begin{array}{cccc}0&0&1&0\\0&0&0&0\\0&0&0&0\\0&0&0&0 \end{array}\right),~
Q_{01}=\left(\begin{array}{cccc}0&0&0&1\\0&0&0&0\\0&0&0&0\\0&0&0&0 \end{array}\right),~
Z=\left(\begin{array}{cccc}0&0&0&0\\0&0&0&0\\0&0&0&1\\0&0&\epsilon&0 \end{array}\right);&
\eea
}}

for $\mu=\lambda$ we get

{\scriptsize{\bea\nonumber
&H=\left(\begin{array}{cccc}\lambda&0&0&0\\0&\lambda&0&0\\0&0&\lambda-1&0\\0&0&0&\lambda-1\end{array}\right),~
Q_{10}=\left(\begin{array}{cccc}0&0&1&0\\0&0&0&p\\0&0&0&0\\0&0&0&0 \end{array}\right),~
Q_{01}=\left(\begin{array}{cccc}0&0&0&1\\0&0&\epsilon p&0\\0&0&0&0\\0&0&0&0 \end{array}\right),~
Z=\left(\begin{array}{cccc}0&q&0&0\\\epsilon p^2q&0&0&0\\0&0&0&1-pq\\0&0&\epsilon(1-pq)&0 \end{array}\right).&
\eea
}}

{\bf Superalgebra $S21_{y}$:}
{\scriptsize{\bea\nonumber
&H=\left(\begin{array}{cccc}\lambda&0&0&0\\0&\lambda-1-y&0&0\\0&0&\lambda-1&0\\0&0&0&\mu\end{array}\right),~
Q_{10}=\left(\begin{array}{cccc}0&0&1&0\\0&0&0&0\\0&0&0&0\\0&0&0&0 \end{array}\right),~
Q_{01}=\left(\begin{array}{cccc}0&0&0&0\\0&0&0&0\\0&1&0&0\\0&0&0&0 \end{array}\right),~
Z=\left(\begin{array}{cccc}0&1&0&0\\0&0&0&0\\0&0&0&0\\0&0&0&0 \end{array}\right);&
\eea
}}

for $\mu=\lambda-y-2$ we get

{\scriptsize{\bea\nonumber
&H=\left(\begin{array}{cccc}\lambda&0&0&0\\0&\lambda-1-y&0&0\\0&0&\lambda-1&0\\0&0&0&\lambda-2-y\end{array}\right),~
Q_{10}=\left(\begin{array}{cccc}0&0&1&0\\0&0&0&p\\0&0&0&0\\0&0&0&0 \end{array}\right),~
Q_{01}=\left(\begin{array}{cccc}0&0&0&0\\0&0&0&0\\0&1&0&0\\0&0&0&0 \end{array}\right),~
Z=\left(\begin{array}{cccc}0&1&0&0\\0&0&0&0\\0&0&0&-p\\0&0&0&0 \end{array}\right);&
\eea
}}

for $\mu=\lambda+y$ we get
{\scriptsize{\bea\nonumber
&H=\left(\begin{array}{cccc}\lambda&0&0&0\\0&\lambda-1-y&0&0\\0&0&\lambda-1&0\\0&0&0&\lambda+y\end{array}\right),~
Q_{10}=\left(\begin{array}{cccc}0&0&1&0\\0&0&0&0\\0&0&0&0\\0&0&0&0 \end{array}\right),~
Q_{01}=\left(\begin{array}{cccc}0&0&0&0\\0&0&0&0\\0&1&0&0\\-p&0&0&0 \end{array}\right),~
Z=\left(\begin{array}{cccc}0&1&0&0\\0&0&0&0\\0&0&0&0\\0&0&p&0 \end{array}\right);&
\eea
}}

for $\mu=\lambda-y$ we get
{\scriptsize{\bea\nonumber
&H=\left(\begin{array}{cccc}\lambda&0&0&0\\0&\lambda-1-y&0&0\\0&0&\lambda-1&0\\0&0&0&\lambda-y\end{array}\right),~
Q_{10}=\left(\begin{array}{cccc}0&0&1&0\\0&0&0&0\\0&0&0&0\\0&0&0&0 \end{array}\right),~
Q_{01}=\left(\begin{array}{cccc}0&0&0&p\\0&0&0&0\\0&1&0&0\\0&0&0&0 \end{array}\right),~
Z=\left(\begin{array}{cccc}0&1&0&0\\0&0&0&0\\0&0&0&0\\0&0&0&0 \end{array}\right).&
\eea
}}

\par

~
\par
~\par
\par

\par {\Large{\bf Acknowledgments}}
{}~\par{}~\par

 The work was supported by CNPq (PQ grant 308095/2017-0).

{}~\par{}~\par

\par {\Large{\bf Data Availability Statement}}
{}~\par{}~\par

 The data that supports the finding of this study are available within the article.

\end{document}